\documentclass[onecolumn,prd,aps,floats,floatfix,nofootinbib,11pt]{revtex4}
\usepackage{verbatim,graphicx,amssymb,amsbsy,bm,amsmath,rotating,epsfig}
\usepackage{psfrag}
\usepackage{hyperref}
\usepackage{fontenc}
\usepackage[latin1]{inputenc}
\usepackage{pstricks,pst-node,pst-text,pst-3d}
\usepackage{textcomp}
\newcommand{\be}{\begin{equation}}
\newcommand{\ee}{\end{equation}}
\newcommand{\bea}{\begin{eqnarray}}
\newcommand{\eea}{\end{eqnarray}}

\newcommand{\vx}{\ensuremath{\vec{x}}}

\begin{document}
\title{Dark Energy is the Cosmological Quantum Vacuum Energy of Light Particles.\\
The Axion and the Lightest Neutrino}
\author{\bf H\'ector J. de Vega $^{(+)}$} 
\author{\bf Norma G. Sanchez $^{(a)}$}
\email{Norma.Sanchez@obspm.fr} 
\affiliation{$^{(+)}$ CNRS LPTHE, Sorbonne Universit\'e, Universit\'e Pierre et Marie Curie UPMC, Paris, Cedex 05, France. \\
$^{(a)}$ CNRS  PSL-Observatoire de Paris, Sorbonne Universit\'e  \\ 
and The Chalonge - de Vega  International School Center, Paris, France.}

\date{\today}
\begin{abstract}
We uncover the general mechanism and {\it the nature} of the today dark energy (DE). This is only based on well known quantum physics and cosmology. We show that the observed DE today originates from the cosmological  quantum {\it vacuum of light particles} which provides a continuous energy distribution able to reproduce the data. Bosons give positive contributions to the DE while fermions yield negative contributions. As usual in field theory, ultraviolet divergences are subtracted from the physical quantities. The subtractions respect the symmetries of the theory and we normalize the physical quantities to be zero for the Minkowski vacuum.  The resulting finite contributions to the energy density and the pressure  
from the quantum vacuum {\it grow} as $ \log a(t) $ where $ a(t) $ is the scale factor,
while the particle contributions dilute as $ 1/a^3(t) $, as it must be for massive particles.
 We find the explicit today dark energy equation of state $ P = w(z) \; {\cal H} $: It turns to be slightly $ w(z) < - 1 $ with $ w(z) $ 
asymptotically reaching the value $ - 1 $ {\it from below}. 
A  scalar particle can produce the observed dark energy through its quantum
cosmological vacuum {\it provided}: {\it (i)} Its mass is of the order of $10^{-3}$ eV = 1 meV,
{\it (ii)}  It is very weakly coupled, and {\it (iii)} it is stable on the time scale of the age of the universe.
The {\it axion vacuum} thus appears as a natural candidate. The neutrino vacuum
(especially the lightest mass eigenstate) can give negative contributions to the dark energy.
We find that $ w(z = 0 ) $ is slightly below $ - 1 $ 
by an amount ranging from $ (-1.5 \times 10^{-3})$ to $ (-8 \times 10^{-3}) $  
and {\it we predict} the {\it axion mass} be in the range between  
4  and  5  ${\rm meV}$ .
We find that the universe will expand in the future {\it faster} than the de Sitter universe, as
an exponential in the square of the cosmic time.
Dark energy today arises from the  quantum {\it vacuum of light particles} in FRW cosmological space-time
in an analogous way to the Casimir vacuum effect of quantum fields in Minkowski space-time with non-trivial boundary conditions.

\bigskip

(+) passed away \url{https://chalonge-devega.fr/HdeV.html}

(a): \url{https://chalonge-devega.fr/sanchez/}
\end{abstract}

\maketitle
\tableofcontents

\section{Introduction and Results}

Since the discovery of the dark energy in the present universe \cite{descu}, \cite{schmidt}, an intense
observational activity has improved our knowledge about it \cite{scp}, \cite {mas}, \cite {DES}, and more activity is expected to provide new data and understanding eg.  \cite{Euclid},  \cite{LSST}. Many different approachs and models have been proposed to explain the dark energy \cite{quinta}, \cite{ mnudt}, \cite{revde}, \cite{albrecht}, \cite{frieman}. For reviews and approachs on the dark energy see for example Refs. \cite {revde}, \cite{quinta}, \cite {mnudt}, \cite{albrecht, frieman}. 

As is by now well known, let us mention that there exist current discordances between different cosmological probes, mainly the discrepancy in the value of the Hubble constant $H_0: 5.0 \; \sigma$ between early universe indirect $H_0$ determinations and late universe direct measurements of $H_0$,  and other stresses and anomalies of lower statistical significance, which are interesting in their own but are not the subject of this paper,  see for example Ref \cite{abdalla2022snowmass} and references therein.
As is well known too, there exist theoretical discordances too, as the fine adjustment of the cosmological constant $\Lambda$, see for example Refs. \cite{albrecht}, \cite{frieman} and references therein. Clarification to this problem have been provided recently  \cite{NSPRD2021}, \cite{NSIJMPA2019}: The huge difference between the observed value of $\Lambda$ today and the particle physics evaluated value $\Lambda_Q$ is correct and must be physically like that, because the two values correspond to the same physical magnitude but to two different vacuum states and cosmic eras: The observed $\Lambda$ value today corresponds to the classical/semiclassical, large and dilute (mostly empty) universe today, consistent with the very low observed $\Lambda$ value, ($10^{-122}$ in Planck units), while the computed value $\Lambda_Q$ ($10^{+122}$ in Planck units) corresponds to the small, highly dense and energetic quantum gravity universe in its far (trans-Planckian) past, and this is consistent with its extremely high, trans-Planckian, value. The two values are classical-quantum duals of each other in the sense of the classical-quantum (wave-particle) duality including gravity, and independently agree with a path integral gravity derivation  \cite{NSIJMPD2019}, \cite{NSIJMPA2019}, \cite{NSPRD2021}.

\bigskip

{\it In this paper} we study the cosmological Quantum Field Theory (QFT) vacuum as dark energy, within a fundamental analytic framework with explicit and analytic results: eg the derivation of the dark energy equation of state and the future evolution of the universe. Moreover, from these results we extract too the implications and determination of the particles contributing to dark energy and compute their masses.

\bigskip

{\it We show} that the dark energy present today in the universe originates
from the cosmological quantum vacuum of light  particles in the meV mass scale. This is a vacuum effect
which {\it unavoidably} appears when quantum fields evolve in a cosmological space-time.
That is,  dark energy today is generated by a mechanism based
on well known quantum physics and cosmology. Bosons yield positive contributions
to the dark energy while fermions give negative contributions. 

{\it We find} that the scale of the contributions to the dark energy is of the order of 
\be 
\frac{M^4}{2 \, (4 \pi)^2} \; \log  z_{\,\rm dec} \; ,
\ee
where $ M $ is the particle mass and $ z_{\rm dec} $ is the redshift when it decoupled
from the early universe plasma.

\bigskip

Generally speaking, the energy of a quantum field is the sum of the vacuum
contribution plus particle contributions.
It is known that the vacuum energy of a quantum field dissipates into particles
when the field evolves coupled to other fields or to itself \cite{nosmink}-\cite{nosb2}.
Dissipation into fermions is reduced by Pauli blocking \cite{nosf,bhp}. 
Electrons, protons and photons are coupled to photons and therefore, 
their vacuum energy dissipates through photon production well before recombination,
that is, when the temperature of the universe was $ 1 $ MeV or more.
Unstable particles cannot produce long-lasting vacuum effects. 
Only a very weakly coupled stable particle can produce
a vacuum energy contribution lasting for times of the order
of the age of the universe, that is, a vacuum energy contribution measurable today.

\bigskip

Since the dark energy is known to be positive, bosons must dominate the cosmological
vacuum energy. The scale of the boson mass must be in the meV range because the observed dark energy
density has the value \cite{slac}
\be\label{datoI} 
\rho_{\Lambda} \;= \; \Omega_{\Lambda} \; \rho_c \;= \;(2.39 \;  {\rm meV})^4 \quad , 
\quad 1 \;  {\rm meV} \;=\; 10^{-3}\;   {\rm eV}
\;\;  .
\ee
Spontaneous symmetry breaking of continuous symmetries is a natural way
to produce massless scalars (Goldstone bosons) in particle physics.
Furthermore, a slight violation of the corresponding symmetry can give
a small mass to such scalar particle. Axions, majorons and familons
have been proposed on these grounds \cite{axi0}, \cite{invi},  \cite{majo}. 

In addition, the lightest neutrino can give a negative contribution to the dark energy.

Neutrinos are by now very well motivated particles from the point of view of  particle physics,  cosmology and astrophysics, eg \cite{dod},  \cite{kt}, \cite {gorbunov}.  For Majorana type neutrinos, neutrinos and antineutrinos coincide while for Dirac neutrinos, neutrinos and antineutrinos are distinct. It is not yet clear whether neutrinos are of Majorana or Dirac type and in this paper we discuss the implications for dark energy of {\it both} of them. Interestingly enough light meV neutrinos and the meV axion do appear here {\it as a consequence} of our results for the dark energy computed from first principles. For constraints on other types of neutrinos and other relativistic species or"dark radiation" see for example \cite{archidiacono} and references therein.

\medskip

Neutrinos in the universe are known to be free for temperatures $ T \lesssim 1 $ Mev 
which correspond to redshifts $ z \lesssim 6 \times 10^9 $ \cite{dod},  \cite{kt}, \cite {gorbunov} 
That is, we can describe their evolution as free fermions in the cosmological FRW
universe. 

Axions with masses $ M \sim 1$ meV are free for temperatures $ T \lesssim 10^6$ GeV which
correspond to redshifts $ z \lesssim 10^{19}$ \cite{axi}. They can be considered as free scalars
in the cosmological FRW universe. Both, the axion and neutrino decoupling happens 
during the radiation dominated era.
Before decoupling, the non-negligeable interaction of the corresponding particles made
dissipation important and therefore the vacuum energy can only become significant after decoupling.
Therefore, we can restrict ourselves to study
the free quantum field evolution in the cosmological space-time after decoupling.

\begin{itemize}

\item{{\it We investigate} the evolution of scalars and fermions as an initial value problem (Cauchy problem) for the
corresponding quantum fields on a cosmological space-time.}

\item{ {\it We find} that the initial
temperature has a negligible effect on the vacuum energy for late times.}

\item{Both axions and neutrinos can lead to vacuum effects lasting cosmological
time scales. Any of the two heavier neutrino mass eigenstates $ \nu_2 $ and $ \nu_3 $
would produce a large negative dark energy in the $ (50 \; {\rm meV})^4 $  range.  Hence:

{\it (i)} either the heavier neutrinos $ \nu_2 $ and $ \nu_3 $ annihilate with their
respective anti-neutrinos in a time scale of the age of the universe, 

{\it (ii)} or a stable scalar particle with mass in the $ \gtrsim 50 $ meV range must be present 
in order to reproduce the observed value of the dark energy Eq.(\ref{datoI}).

However, {\it we find} in this paper that the possibility {\it(ii)} is inconsistent with the observed dark energy equation of state.}
\end{itemize}
An effective four fermions interaction with strength characterized by $ M'^{\,-2} $ where
$ M' $ is a mass scale can make the heavier neutrinos unstable. The mass scale $ M' $ should be
$ M' \lesssim 1$ MeV or  $ M' \lesssim 10$ MeV for the direct and inverse neutrino
mass hierarchies.  

As shown in Section VIII  the lightest meV neutrino remains the only neutrino contribution to the dark energy. The heavier neutrinos dissipate at the time of the age of the universe. 

As shown in Section VII,  the meV axion lifetime to decay into photons is much longer than the age of the universe. Dissipation of the energy in the cosmological quantum axion vacuum takes longer than the age of the universe too.

These results are unified in Section IX  with both light meV particles contributing to dark energy together: meV axions and meV light neutrinos,  Table 1  summarizes their contributions, together with the computed  equation of state.

On the other hand, let us mention that a global analysis of cosmological constraints on decaying axion-like particles  (ALPs) performed recently Ref \cite{BalazsALPs2022}  shows that ALPs are stable on cosmological time scales unless they would be heavy enough  with masses $ > 300$ keV. This is an independent confirmation  that $10^{-3}$ eV axions as shown in this paper are safely enough stable to be considered as the source of dark energy. Previously, ALPs have been proposed among other proposals, to be constituents of the cosmological energy density  i.e. Ref \cite{FerreiraALPs2015}.

\bigskip

In conformal time $(\eta)$, the scalar and fermion fields rescaled by the scale factor $ a (\eta) $
turn out to obey equations of motion similar to those in Minkowski space-time but with time-dependent
masses
\bea\label{ecmovI}
&& \chi'' \;-\; \nabla^2 \chi \;+\;  \left[\;M^2 \; a^2(\eta)\;-\;
\frac{a''(\eta)}{a(\eta)} \;\right] \chi({\vec x},\eta)\; = \;0\; , \cr \cr
&& \left[\;i \; {\not\!{\partial}}\;- \;m \; a(\eta) \;\right]\, \psi({\vec x},\eta)\, =\, 0 \; .
\eea
Here, $ \chi $ and  $ \psi $ are, respectively, rescaled scalar and fermion fields,
$  \nabla^2 $ is the usual flat space Laplacian and $ i {\not\!{\partial}}$  is the usual 
Dirac differential operator in Minkowski space-time in terms of flat
space-time Dirac matrices.

\bigskip

There are two widely separate scales in the field evolution in  cosmological space-times:
\begin{itemize}
\item{The fast scale is the microscopic quantum evolution scale,\\
typically $ \sim 1/M \sim 1/m $,
where $ M $ and $ m $ are the scalar and fermion masses respectively.}
\item{The slow scale is the Hubble scale $ 1/H $ of the universe expansion.\\
When $ M \sim m \gg H , \; M^2 \gg a''(\eta)/a^3(\eta) $, and hence the scale factor
can be considered as constant.} \item{Therefore, 
the cosmological quantum field evolution for the fields $ \chi $ and $ \psi $ is just the Minkowski evolution 
with effective masses $( M^2\, a^2 )$ and $ (m\,a) $, respectively, as seen from Eq.(\ref{ecmovI}).}
\end{itemize}
Energy density, pressure, and field density express in field theory as products
of the field operators and their derivatives at equal space-time points.
Such expressions are ultraviolet divergent and need to be subtracted. 
The subtractions respect the symmetries of the theory
and we normalize them such that the physical quantities are zero for the vacuum
in Minkowski space-time. The finite resulting quantities grow as $ \log a(\eta) $.
This is analogous to the high-energy growth of renormalized one-loop Feynman graphs.

That is, the energy density and the pressure get contributions
from the quantum vacuum that grow as $ \log a(\eta) $ while the particle
contributions dilute as $ 1/a^3(\eta) $, as it must be for massive particles.

\bigskip

{\it We obtain} for the vacuum energy density and pressure of scalar and fermion fields
with mass $ M $ and $ m $, respectively the following results:
\be\label{hpI}
<{\cal H}>(\eta) \buildrel{ a(\eta) \;\gg \;a_{\rm dcs}, \; a_{\rm dcf}}\over=  
\frac{M^4}{2 \, (4 \, \pi)^2}\left[\; \log a(\eta)+ b_S 
 -\frac14 \;\right] - \frac{m^4}{(4 \, \pi)^2}\; \mathcal{N} \; \left[\; \log a(\eta) + b_F -\frac14 \;\right] ,
 \ee
 \be
 <P>(\eta) \buildrel{ a(\eta) \;\gg \;a_{\rm dcs}, \; a_{\rm dcf}}\over = 
- \frac{M^4}{2 \, (4 \, \pi)^2}\;\left[\; \log a(\eta) + b_S + \frac1{12} \; \right]
+ \frac{m^4}{(4 \, \pi)^2}\; \mathcal{N} \;\left[ \;\log a(\eta) + b_F + \frac1{12} \;\right] 
\ee
where  $ b_S $ and  $ b_F $ take into account the initial values of the scale factor $ a_{\rm dcs} $ and $ a_{\rm dcf} $
(at the decoupling time) of the scalars and fermions, respectively. 
$ \mathcal{N} = 1 $ for Majorana fermions and $ \mathcal{N} = 2 $ for Dirac fermions. 

\bigskip

Therefore,  {\it we obtain} for  the equation of state the {\it explicit} expression:

\be\label{asiIn1}
w(\eta)\; \equiv \;\frac{ <P>(\eta)}{ <{\cal H}>(\eta)} 
\buildrel{ a(\eta) \;\gg \;a_{\rm dcs},\;  a_{\rm dcf}}\over=  - 1\, -\, \frac13 \, \left[\; \log a(\eta)\,-\,\frac14 \,
+\,\frac{b_S - (\,2 \; \mathcal{N} \; m^4/{M^4}\,)\; b_F }{1 - (\,2 \;\mathcal{N} \; m^4 / {M^4}\,)}\;\right]^{-1}\;
\ee
That is,  we find $ w(\eta) < - 1 $ with  $ w(\eta) $ asymptotically reaching the value
$ - 1 $ {\it from below}. 

\medskip

It is convenient to express the scale factor in terms of the redshift.
Taking into account that  $b_S $ and  $ b_F $ contain the initial values
of the scale factor yields,
\be\label{azI}
a(\eta) \; e^{b_S}\; = \;\frac{1\; +\;z_S}{1+z} \quad ,  \qquad 
a(\eta) \; e^{b_F} \;= \;\frac{1\; +\;z_F}{1+z} \; ,
\ee
where $ z_S $ ($ z_F $) is the redshift when the scalar (fermion) field decoupled.
For neutrinos, $ z_F \sim 6 \times 10^9 $, while
 for axions with mass $ \sim 1 $ meV, $ z_S \sim 2.2 \times 10^{18} $. 

{\it We find} from Eqs.(\ref{hpI}) and (\ref{azI}),
\bea\label{hpzI}
&& <{\cal H}>(z) = \frac1{2 \, (4 \, \pi)^2}\left\{ M^4 \; \log z_S 
-2 \; \mathcal{N} \; m^4  \; \log z_F -( M^4-2 \; \mathcal{N} \; m^4)\left[ \; \log (1 + z) + \frac14 \;\right]  
\right\}\\ \cr
&& <P>(z) = -\frac1{2 \, (4 \, \pi)^2}\left\{ M^4 \; \log z_S 
-2 \; \mathcal{N} \; m^4  \; \log z_F  -( M^4-2 \; \mathcal{N} \; m^4)\left[\; \log (1 + z) -\frac1{12} \;\right]  
\right\} \nonumber
\eea

where we used that $ z_S \gg 1, \; z_F  \gg 1 $.

\medskip

We identify the vacuum energy density today 
$ <{\cal H}>(z = 0) $ with the observed dark energy $ \rho_{\Lambda} $.
We can then write Eqs.(\ref{hpI}),  (\ref{asiIn1}) and (\ref{hpzI}) as:
\bea
&&\rho_{\Lambda}\; = \;\frac1{2 \, (4 \, \pi)^2}\left[\; M^4 \left( \log z_S -\frac14 \right) 
-2 \; \mathcal{N} \;  m^4 \left(  \log z_F  -\frac14 \right) \;\right] \; , \label{3nuI} \\ \cr
&&<{\cal H}>(\eta) \buildrel{ a(\eta) \;\gg \; a_{\rm dcs},\; a_{\rm dcf}}\over=
  \rho_{\Lambda} \left[\; 1 \;+ \;\beta_{\mathcal{N}} \;  \log \frac{a(\eta)}{a_0} \;\right] 
\label{rhoetaI} \; , \cr \cr
&& w(\eta) \;+;\ 1 \buildrel{ a(\eta)\; \gg \;a_{\rm dcs}, \; a_{\rm dcf}}\over= - \;
\frac{ \left(M^4\;- \;2 \, \;\mathcal{N} \; m^4\right)}{6 \, (4 \, \pi)^2 \;\rho_{\Lambda} 
\left[\; 1\; + \;\beta_{\mathcal{N}} \;  \log \frac{a(\eta)}{a_0} \;\right]} \; 
\eea 
where $ a_0 $ is the scale factor today and 
\be\label{defbetaI}
\beta_{\mathcal{N}} \;= \; \frac{\displaystyle \left(1  - \frac{2 \; \mathcal{N} \; m^4}{M^4}\right)}{\displaystyle \log z_S \;-\; \frac14 \;
- \;\left (2\; \frac{\; \mathcal{N} \; m^4}{M^4}\right) \left[\; \log z_F - \frac14\;\right] \;} \; .
\ee
That is, the vacuum energy density at late times after decoupling {\bf grows} as the logarithm
of the scale factor and the equation of state asymptotically approaches $ - 1 $ {\it from below}.

\medskip

 The {\it equation of state} as a {\it function of  z} takes the form:
\be\label{wmzI}
w(z)\; +\; 1 \;= \; - \;\frac13 \; \frac{\displaystyle \left(1  - \frac{2 \; \mathcal{N} \; m^4} {M^4} \right)}{\displaystyle \log z_S - \left(\frac{2 \; \mathcal{N} \; m^4} {M^4}\right)
 \log z_F  - \left(1  - \frac{2 \; \mathcal{N} \; m^4} {M^4}\right)\left[\; \log(1 + z) + \frac14 \;\right] }  
\ee
For $ z = 0 $, it becomes today:
\be\label{ecshoyI}
w(0)\;+ \;1 = -\; \frac13 \; \frac{\displaystyle \left(1  - \frac{2 \; \mathcal{N} \; m^4} {M^4}\right) }{\displaystyle
 \log z_S - \frac14 - \left(\frac{2 \; \mathcal{N} \; m^4}{M^4}\right) \left[\;\log z_F -\frac14 \; \right]} = 
 -\; \frac{1}{6 \,(4 \,\pi)^2 \,\rho_{\Lambda}}  \left(M^4- 2 \,\mathcal{N} \; m^4 \right) \; 
\ee
The scalar and fermion masses are constrained by the value of the dark energy today
Eq.(\ref{datoI}). This gives the positivity requirement:
$$
M \;>\; \left( 2 \; \mathcal{N} \right)^{\frac14} \; m \; ,
$$
as well as the expression for the mass of the scalar particle:

\be\label{cotagI}
M \; = \; \displaystyle{ \frac{10.1 \; {\rm meV}}{\left( \;\log z_S \;- \; \frac14 \;\right)^{ \, \frac14}}\,
\left[\; 1 \,+ \,\mathcal{N} \; \left(\; \frac{m}{3.90 \; {\rm meV}} \; \right)^{\! 4} \; \right]^{\,\frac14}} \;\;
\ee 
The neutrino contribution to the dark energy 
can be ignored when  $ m \ll 1 $ meV and when the vacuum neutrino contribution dissipates 
in the time scale of the age of the universe as mentioned before.
The mass of the lightest neutrino is not yet known (only neutrino mass
differences are known). We will consider that the lightest neutrino mass is either
$ m = 3.2 $ meV \cite{masanu}, \cite{masanu2} or zero \cite{chica}.

More specifically, we set $ z_S \sim 2.2 \times 10^{18} $ assuming the scalar field to be an axion 
with mass $ \sim 1 $ meV in Eqs.(\ref{ecshoyI}).  (\ref{cotagI}). 
\begin{itemize}
\item{We therefore  obtain for the axion mass $M$ and for the equation of state today the following values:
\bea\label{cotaxiI}
&&3.96 \; {\rm meV}\; < \;M \; < \;4.66 \; {\rm meV} \; , \cr \cr
&& -\;0.00794\; < \; w(0) \;+ \;1 \;< \; -\;0.00156 \; 
\eea
The left and right ends of the intervals in Eq.(\ref{cotaxiI})
correspond respectively to no neutrino contribution, and to the lightest neutrino  contribution as
a Dirac fermion with mass $ m = 3.2 $ meV.}

\item{ We see that $ w(0) $ is slightly below $ - 1 $ by an amount ranging
from $ (-1.5 \times 10^{-3} )$ to $( -8  \times 10^{-3} )$, while the axion mass results 
{\it between $ 4 $ and $ 5 $} meV which is within the range of axion masses allowed by astrophysical 
and cosmological constraints, eg.  \cite{axiM}.

If the scalar particle is not the axion, the value of $ z_S \gg 1 $ will depend on the
dynamics of such scalar particle.}
\item{ {\it In general}, we express the contribution of the quantum vacuum of light particles
to the dark energy and pressure in terms of {\it two} parameters: the particle masses and the redshifts
when they decoupled. There is also a dependence on the number of states per particle
(1 for a scalar, $ 2 \, \mathcal{N} $ for a fermion).}

\item{  We uncover in this paper the general mechanism producing the dark energy today. This mechanism
is only based on well known quantum physics and cosmology. 
The observed dark energy in the universe today appears as a {\it quantum vacuum effect} only due to the  (classical) cosmological 
space-time expansion. That is to say, dark energy in the present universe is a {\it
semiclassical gravity effect}.}

\item{ The dark energy arises for a quantum field in the cosmological context
in an  analogous way the {\it Casimir effect} arises for a quantum field in Minkowski space-time with non-trivial boundary conditions in space.}
 
\item{All physical (finite) results are independent of any energy cutoff 
as well as of the regularization method used.}

\item {{\it We obtain and solve} in this paper the  {\it self- consistent} Einstein-Friedmann equation
for the scale factor when the dark energy dominates and the universe
expansion accelerates.
The growth of the energy density Eq.(\ref{hpI}) as
the logarithm of the scale factor implies an expansion  faster than in de Sitter space-time.
More precisely, we find that the Universe will reach in  the future an asymptotic phase where it expands exponentially as
\be\label{afuteI}
a(t)\buildrel{ H_0\,  t \; \gtrsim \; 1}\over\simeq a\,({\rm today}) \; 
\exp{[\;c_1 \, H_0 \, t \, + \, c_2 \; (H_0 \; t)^2\;]} 
\ee
where 
\bea\label{bcI}
&& c_1 \;\equiv \;\sqrt{\Omega_{\Lambda}}\; = \; 0.87 \quad , \quad
 0.00452\; <\; c_2 \;< \;0.00872 \; ,
\eea
and $ H_0 $ stands for the Hubble parameter today.
The left and right ends of the interval for $ c_2 $ in Eq.(\ref{bcI})
correspond respectively to the no neutrino contribution and to the lightest neutrino contribution as 
a Dirac fermion with mass $ m = 3.2 $ meV.} 

\item{Notice that the time scale of the accelerated expansion is huge $ \sim 1 \,/ \, H_0  = 13.4$ Gyr.
In the exponent of  Eq.(\ref{afuteI}) the quadratic term dominates over the linear term 
by a time $ t \,\sim \, 100\,/\,H_0$ to  $200 \,/\, H_0$.

In this accelerated universe, we see from the Friedman equation and Eq.(\ref{hpI}), 
that the Hubble radius $ 1/H $  {\it decreases} with time as $ 1\, /\,[ \,H_0 \; \sqrt{\log a(t)}\,] $.}

\end{itemize}

 This paper is organized as follows: In Sections II and III we review the dynamics of 
scalar and fermion fields on cosmological space-times, respectively. 
In Section IV we discus the quantum cosmological vacuum, the two point functions, et compute the main physical quantities from them.
In Section V we find the vacuum energy density, 
pressure and the equation of state for late times and in Section VI we discuss their quantum nature.
In Section VII we find dark energy as a result of the cosmological quantum vacuum contributions from meV light particles and their properties, masses and stability, axions are treated in this Section. In Section VIII  we compute and analyze the neutrino contributions to dark energy: the lightest neutrino remains the only contribution. In Section IX we unify these results with both light meV particles together. We obtain the future self-consistent evolution of the universe in Section X.
We discuss relevant related issues in Section XI, and we present our conclusions in Section XII. 
An Appendix is devoted to the equivalence between different regularization methods.

\section{Scalar Fields in Cosmological Space-Times}\label{esca}

We consider a massive neutral scalar field $ \varphi $ 
in a FRW geometry defined by the invariant distance
\be\label{FRW}
ds^2 \;= \;dt^2\;-\;a^2(t) \; d\vec{x}^2  \; .
\ee
The Lagrangian density is taken to be
\be
\mathcal{L}\; = \;\frac12 \;  \sqrt{-g} \; \Bigg[\; \dot{\varphi}^2 \;-\;
\left(\frac{\vec{\nabla}\varphi}{a}\right)^2 - \;M^2 \; \varphi^2 \;\Bigg] \label{lagra} \; .
\ee
It is convenient to use the conformal time $ \eta $
$$
\eta \;= \;\int \frac{dt}{a(t)}\; ,
$$
and the conformally rescaled field $ \chi(\vx,\eta) $,
\be\label{evcon}
\chi(\vx,\eta)\; \equiv \; a(t) \; \varphi(\vx,t) \; .
\ee
The action (after discarding surface terms that do not affect the equations of motion) reads:
\be\label{confoaction}
A\Big(\chi,\delta\Big) \;=\;\frac12 \; \int d^3x \;  d\eta \;
\Bigg[\; {\chi'}^2 - (\nabla \chi)^2 - \mathcal{M}^2(\eta) \; \chi^2 \; \Bigg]
\ee
\noindent where primes denote derivatives with respect to the conformal
time $ \eta $ and where
\be
\mathcal{M}^2(\eta) \; =\;  M^2 \; a^2(\eta) \;- \;\frac{a''(\eta)}{a(\eta)} 
\; , \label{massdelta}
\ee
plays the role of an effective mass squared.
Therefore, the rescaled field $ \chi(\vx,\eta) $ obeys the equation of motion,
\be\label{ecmovc}
\chi'' \;- \; \nabla^2 \chi \;+ \;\mathcal{M}^2(\eta) \; \chi \;= \;0\; .
\ee
The evolution of  $ \chi(\vx,\eta) $ is like that of a scalar field in Minkowski space-time
with a time-dependent mass squared $ \mathcal{M}^2(\eta) $.

The solution for the field $ \varphi(\vx,t) $ can be Fourier expanded as follows,
\be\label{dce}
\varphi(\vx,\eta) \;= \;\frac1{a(\eta)} \int  \frac{d^3k}{(2\,\pi)^3 \; 2 \; E_0}\; \left[\;
a_{\vec{k}} \;\; \phi_k\;(\eta) \;  e^{i \vec{k}\cdot \vec{x}} \; + \;  
a_{\vec{k}}^{\dagger} \;\; \phi_k^*\;(\eta) \;  e^{-i \vec{k}\cdot \vec{x}}
\;\right]
\ee
where 
$$ 
E_0 \;\equiv \;\sqrt{\;k^2 \;+\;  \mathcal{M}^2_i \;} \; ,
$$
and $ \mathcal{M}_i $ is the effective mass $ \mathcal{M}(\eta) $ at the decoupling
time (initial time) for the scalar field evolution.
The mode functions $ \phi_k(\eta) $ obey the evolution equations,
\be\label{ecfunm}
\left[\; \frac{d^2}{d \eta^2 }\; +\; k^2 \; + \;
 M^2 \; a^2(\eta)\;-\;
\frac{a''(\eta)}{a(\eta)} \;\right] \phi_k(\eta)\; = \;0 \; .
\ee
We choose the initial state as the vacuum state which is here (at decoupling) a thermal
equilibrium state at temperature $ T $. However, as we see below [Eq.(\ref{Tcorr})], 
the effect of the initial
temperature on the vacuum energy is negligible for late times.
The Fock vacuum state $ |0> $ is annihilated by the operators $ a_{\vec k} $.
Therefore, we have as initial conditions for the mode functions,
\be\label{ciB}
\phi_k(0) \;= \;1 \quad , \quad \phi'_k(0) \; = \;- \;i \; E_0 \; .
\ee
These initial conditions describe the Bunch-Davies vacuum when they
are applied at asymptotically earlier times in the past ($ \eta \to -\infty $) \cite{BD}, \cite{anom}.
See the discussion in Sec. \ref{discu} here below.

The time-dependent creation and annihilation operators obey the canonical commutation rules,
$$
\left[\;a_{\vec{k}}\;,\; a_{\vec{k}'}^{\dagger} \;\right] \;= \;2 \; E_0 \; (2\,\pi)^3 \;
\delta(\vec{k}-\vec{k'}) \; .
$$
The energy-momentum tensor for a scalar field is given by \cite{BD},
\be\label{tmne}
T_{\mu \nu} \;= \; \partial_\mu \varphi \; \partial_\nu \varphi\;
- \;\frac12 \; g_{\mu \nu} \left[\; \partial_\lambda \varphi \; \partial^\lambda \varphi
\;- \; M^2 \; \varphi^2 \; \right] \; .
\ee
Its expectation value has the fluid form 
$$
<T_{S \; 0}^0> \;\;  = \;\;  < {\cal H}_S >(\eta) 
\quad ,  \quad <T_i^j> \;\;  = \;-\;\;\delta_i^j \, < P_S >(\eta) \;  \; ,
$$
since we consider homogeneous and isotropic quantum states and density matrices.
In conformal time the hamiltonian density and the pressure take the form
\bea\label{hpes}
&&{\cal H}_S(\eta) \; = \; \frac1{2 \;  a^4(\eta)} \left\{\;
\left[\;\chi'({\vec x},\eta) - a(\eta) \;  H(\eta)\; \chi({\vec x},\eta)\,\right]^2
+   (\,\nabla \chi ({\vec x},\eta)\, )^2 +  a^2(\eta) \; M^2 \; \chi^2({\vec x},\eta) 
\;\right\} \; , \cr \cr  
&& {\cal H}_S\; +\;  P_S(\eta) \;= \; \frac1{a^4(\eta)}\left\{
\left[\;\chi'({\vec x},\eta) - a(\eta) \;  H(\eta)\; \chi({\vec x},\eta)\;\right]^2+
\frac13 \;  (\,{\nabla \chi}({\vec x},\eta)\,)^2 \,\right\} \; ,
\eea
where $ H(\eta) $ stands for the Hubble parameter
\be\label{hub}
H(\eta) \,\,\equiv\;\, \frac{d\;\ln a(t)}{dt} \; = \;\frac1{ a^2(\eta)} \frac{d a}{d \eta}\; .
\ee
It is convenient to consider the conformal energy and pressure,
\be\label{epcon}
{\varepsilon}_S(\eta) \;\equiv\;  a^4(\eta) \; <{\cal H}_S>(\eta) \quad ,  \quad 
 p_S(\eta) \;\equiv \; a^4(\eta) \; < P_S >(\eta) \; .
\ee
We find the trace of the energy-momentum tensor from Eqs.(\ref{hpes}),
\be
 a^4(\eta)\left[ \,{\cal H}_S(\eta) - 3 \;  P_S(\eta)\,\right] \;= \;a^2(\eta) \;
M^2 \; \chi^2 - \left[\,(\chi'- a \; h \;\chi)^2 -
(\nabla \chi)^2 -  a^2(\eta) \; M^2 \; \chi^2\,\right] \; .
\ee
Ignoring the bracket term in the right hand side yields the virial theorem.
Although this bracket term is nonzero, its space and time average is zero:
$$
\frac1{\Delta}  \int_{\eta}^{\eta+\Delta} d \eta \int d^3x \left[(\chi'- a \; h \;\chi)^2 -
(\nabla \chi)^2 -  a^2(\eta) \; M^2 \; \chi^2 \right] 
\buildrel{\Delta \gg 1/M}\over= 0 \; .
$$
In addition, this bracket can be neglected for late times as we shall see below. 

Therefore, we have for the expectation values,
\be\label{virB}
{\varepsilon}_S(\eta)  - 3 \, p_S(\eta)\; = \;M^2 \; a^2(\eta) \;  \Sigma_S(\eta) 
-  a^4(\eta) \; V(\eta) 
\ee
where 
\be \label{dfSB}
 \Sigma_S(\eta) \;\equiv\; <\chi^2({\vec x},\eta)>\; = \;
a^2(\eta) \;  <\varphi^2({\vec x},\eta)>  \; ,
\ee
and $ V $ stands for the expectation value of the virial
$$
V(\eta)  \; \equiv \; <(\chi'- a \; h \;\chi)^2 -
({\nabla \chi})^2 - a^2(\eta) \; M^2 \; \chi^2> \; .
$$

\medskip

Using the equations of motion (\ref{ecmovc}) we obtain for the time derivative 
of the energy density Eq.(\ref{epcon}),
\be\label{cetaB}
\frac{d \varepsilon_S}{d\eta} \;= \;\frac12 \; M^2 \; \; \frac{d a^2(\eta)}{d \eta} \, \; \Sigma_S(\eta)
\; - \; a(\eta) \;  H(\eta) \; V(\eta) \; .
\ee
This relation in conformal time implies the usual continuity equation in cosmic time
\be\label{contB}
\frac{d}{dt} <{\cal H}_S>\; + \;\, 3 \; H(\eta) \; \left[\, <{\cal H}_S>\; + \;<P_S> \,\right] \;= \; 0 \; .
\ee
Therefore,  from Eqs.(\ref{virB}) and (\ref{cetaB}) we see that there is only one independent quantity among 
$ \varepsilon_S(\eta), \; p_S(\eta) $ and $ \Sigma_S(\eta) $.

\section{Fermion Fields in Cosmological Space-Times}\label{suno}

The Lagrangian density for fermions is taken to be \cite{anom}
\be
\mathcal{L} \;=\; \sqrt{-g} \;\; 
\overline{\Psi}\,\Big[\;i\,\gamma^\mu \;  \mathcal{D}_\mu \Psi \;-\;m \;\Big]\,\Psi \; . \label{lagrangian}
\ee
\noindent 
The $ \gamma^\mu $ are the curved space-time Dirac $ \gamma $ matrices
and the fermionic covariant derivative is given by
\bea
\mathcal{D}_\mu & = &  \partial_\mu\; +\; \frac{1}{8} \;
[\;\gamma^c\;,\;\gamma^d  _;] \;  V^\nu_c  \; \left(\;D_\mu V_{d \nu}\; \right)
\cr \cr 
D_\mu V_{d \nu} & = & \partial_\mu V_{d \nu}\; -\;\Gamma^\lambda_{\mu
\nu} \;  V_{d \lambda} \nonumber
\eea
\noindent where the vierbein field is defined as
$$
g^{\mu\,\nu} \; = \; V^\mu_a \;  V^\nu_b \;  \eta^{a b} \; ,
$$
\noindent $ \eta_{a b} $ is the Minkowski space-time metric and
the curved space-time  matrices $\gamma^\mu$ are given in terms of
the Minkowski space-time ones $\gamma^a$  by (Greek indices refer to
curved space time coordinates and Latin indices to the local
Minkowski space time coordinates)
$$
\gamma^\mu \; = \; \gamma^a \; V^\mu_a \quad , \quad
\{\gamma^\mu,\gamma^\nu\}\;= \;2 \; g^{\mu \nu}  \; .
$$
In conformal time the vierbeins $ V^\mu_a $ are particularly simple
\be
V^\mu_a \; = \; a(\eta) \; \delta^\mu_a \; ,
\ee
\noindent where $ a(\eta) \equiv a(t(\eta)) $ is the scale factor
as a function of the conformal time and we call $ a(\eta=0) = a_{\rm dc} $.
The Dirac Lagrangian density thus simplifies to the following expression
\be \label{ecdi}
\sqrt{-g} \; \overline{\Psi}\left(\;i \; \gamma^\mu \;  \mathcal{D}_\mu
\Psi \;-\; m \;\right)\Psi \; =\;
a^{\frac32}\;\overline{\Psi} \;  \left[\;i \;
{\not\!{\partial}}\; -\; m  \; a(\eta)\; \right]
\left(a^{\frac{3}{2}}{\Psi}\right)
\ee
\noindent where $ i {\not\!{\partial}}$  is the usual Dirac
differential operator in Minkowski space-time in terms of flat
space time $ \gamma^a $ matrices.

Therefore, the Dirac equation in the FRW geometry is given by 
\be\label{pamd}
\left[\;i \; {\not\!{\partial}}\;-\;m \; a(\eta) \;\right]
\left[\;a^{\frac32}\;{\Psi({\vec x},\eta)}\;\right]\; =\; 0 \; .
\ee
The solution $ \Psi({\vec x},\eta) $ can be expanded in spinor mode
functions as
\be\label{psiex}
\Psi(\vec{x},\eta) \; =\; \frac{1}{a^\frac32 (\eta)} \;  
\sum_{\lambda\,=\,\pm 1} \int \frac{d^3k}{(2\,\pi)^3 \; 2 \; e_0 } \;  
e^{i \vec{k}\cdot \vec{x}} \;  
\left[\;b_{\vec{k},\lambda}\, U_{\lambda}(\vec{k},\eta)\;+\;
d^{\dagger}_{\,-\vec{k},\lambda}\, V_{\lambda}(-\,\vec{k},\eta)\;\right] \; ,
\ee
where 
$$ 
e_0 \;\equiv \;\sqrt{\;k^2 \;+ \;m^2 \; a^2_{\rm dc}} \; ,
$$
and the spinor mode functions $ U,V $ obey the  Dirac equations
\bea
\left[\,i \; \gamma^0 \;  \partial_\eta \;- \;\vec{\gamma}\cdot \vec{k}
 \;- \;m \; a(\eta) \,\right]U_\lambda(\vec{k},\eta) & = & 0 \label{Uspinor} \\
\left[\,i \; \gamma^0 \;  \partial_\eta \;+ \;\vec{\gamma}\cdot \vec{k}\; -\;m \; a(\eta)
\,\right]V_\lambda(\vec{k},\eta) & = & 0 \label{Vspinor} \; .
\eea
The time-independent creation and annihilation operators obey
the canonical anticommutation rules
\bea
&&\left\{b_{\vec{k},\,\lambda}\;,\;b^{\dagger}_{\vec{k'},\,\lambda'}\right\}\;\;=\,2 \; e_0 \; (2\,\pi)^3 \;
\delta(\vec{k}-\vec{k'}) \; \delta_{\lambda \; \lambda'} \; , \cr \cr
&& \left\{d_{\vec{k},\,\lambda}\;,\;d^{\dagger}_{\vec{k'},\lambda'}\right\}\;=\;2 \; e_0 \; (2\,\pi)^3 \;
\delta(\vec{k}-\vec{k'}) \; \delta_{\lambda \; \lambda'} \; .\label{CAR}
\eea
Following the method of Refs. \cite{nosf}, \cite{bhp}, it proves convenient to write
\bea
U_\lambda(\vec{k},\eta) & = & {(e_0 \;+ \;m \; a_{\rm dc})}^{-\frac12} \; 
\left[\,i \; \gamma^0 \;  \partial_\eta \;-\;
\vec{\gamma}\cdot \vec{k}\; + \;m \; a(\eta)
\,\right]f_k(\eta) \; \mathcal{U}_\lambda \label{Us}\\ \cr
V_\lambda(\vec{k},\eta) & = &  {(e_0 \;+ \;m \; a_{\rm dc})}^{-\frac12} \; 
\left[\,i \; \gamma^0 \;  \partial_\eta \;+\;
\vec{\gamma}\cdot \vec{k} \;+\;m \; a( \eta)
\,\right]g_k(\eta) \; \mathcal{V}_\lambda \label{Vs} \; ,
\eea
\noindent $  {(\;e_0 + m \; a_{\rm dc}\;)}^{-\frac12} $ being a normalization factor
and $( \mathcal{U}_\lambda , \; \mathcal{V}_\lambda )$ being
constant spinors \cite{nosf}, \cite{bhp} obeying
\be
\gamma^0 \; \mathcal{U}_\lambda \; = \; \mathcal{U}_\lambda
\label{Up} \qquad , \qquad
\gamma^0 \;  \mathcal{V}_\lambda \; = \; - \;\mathcal{V}_\lambda \qquad , \quad \lambda \;= \;\pm \;1 \; .
\ee
More explicitly,
\bea
&& U_\lambda(\vec{k},\eta) \;= \;{(e_0 \;+\; m \; a_{\rm dc})}^{-\frac12} \; 
\left( \begin{array}{cc}
\left[\,i \; f'_k(\eta) \;+ \;m \; a( \eta) \; f_k(\eta)\,\right]  & 0 \\
  0 & \lambda \; k \; f_k(\eta)    \\
  \end{array} \right) \mathcal{U}_\lambda \;  \; ,  \cr \cr  \cr
&& V_\lambda(-\vec{k},\eta) = {(e_0 \;+ \;m \; a_{\rm dc})}^{-\frac12} \; 
\left(  \begin{array}{cc}
  \lambda \; k \; g_k(\eta) & 0  \\
0 & \left[\,-i \; g'_k(\eta) \;+ \;m \; a( \eta) \; g_k(\eta) \,\right] 
  \end{array} \right)\mathcal{V}_{\lambda}  \;  \; .
\eea
The mode functions $ f_k(\eta), \; g_k(\eta) $ obey then the following equations of motion
\bea \label{eqm}
\left[\;\frac{d^2}{d\eta^2} \;+\;
k^2\;+\;m^2 \; a^2(\eta)\;-i \; m \; a'(\eta)\;\right]f_k(\eta) & = & 0 \\ \cr
\left[\;\frac{d^2}{d\eta^2}\; +\; k^2\;+\;m^2 \; a^2(\eta)\;+\;i \; m \; a'(\eta)\;\right]g_k(\eta)
& = & 0  \label{eqmg}\; .
\eea
We choose the initial state for the fermion field as the vacuum state which is a thermal
equilibrium state at temperature $ T $ for the fermion.
This Fock state $ |0> $ is annihilated by the operators $ b_{\vec{k},\lambda} $
and $ d_{\vec{k},\lambda} $.

Therefore, we have as initial conditions for the mode functions \cite{nosf}, \cite{bhp}
\bea \label{condi}
&& f_k(0)\; = \; 1 \quad ,  \quad f'_k(0) \;= \;-\; i \; e_0  \quad ,  \\ \cr
&& g_k(0) \;= \;1 \quad ,  \quad g'_k(0) \;= \;+ \;i \; e_0 \nonumber
\eea
These initial conditions describe the Bunch-Davies vacuum when they
are applied at asymptotically earlier times in the past ($ \eta \to -\infty $) \cite{BD},  \cite{anom}.
See the discussion in Sec. \ref{discu} below.

Eqs.(\ref{eqm})-(\ref{condi}) imply that
$$
 g_k(\eta) \;=\;  f_k^*(\eta) \; .
$$
That is, we have only one independent and complex mode function.

The scalar products of the spinors  $ U_{\lambda}(\vec{k},\eta),
\;  V_{\lambda}(\vec{k},\eta) $ take the values
\bea
&&U^{\dagger}_{\lambda}\;(\vec{k},\eta) \; U_{\lambda'}(\vec{k},\eta) \;= \;2 \; e_0 \;
\delta_{\lambda \; \lambda'} \cr \cr
&&V^{\dagger}_{\lambda}\;(\vec{k},\eta) \; V_{\lambda'}(\vec{k},\eta) \;= \;2 \; e_0 \;
\delta_{\lambda \; \lambda'}
\eea
As a consequence, the mode functions obey the relation \cite{nosf}, \cite{bhp}
$$
|f'_k(\eta)|^2 - i \, m \; a(\eta) \; [f_k(\eta) \; {f'}_k^*(\eta)
- f'_k(\eta) \;  f_k^*(\eta)] + [k^2 + m^2 \; a^2(\eta)] \; |f_k(\eta)|^2
= 2 \; e_0 (e_0 + m \; a_{\rm dc}) \; ,
$$
which provides a conserved quantity. 

The energy momentum tensor for a spin $1/2$ field is given by \cite{BD}
\be \label{tmunu}
T^F_{\mu \nu} \;=\; \frac{i}{2}\,\left[\,\overline{\Psi} \gamma_{(\mu}
\stackrel{\leftrightarrow}{\mathcal{D}}_{\nu)}\Psi \,\right] ,
\ee
\noindent and its expectation value has the fluid form 
$$
<{T_F}_0^{\,0}> \; \; = \;\;  <{\cal H}_F>(\eta) \quad ,  \quad <{T_F}_i^{\,j}> \; \; = \;- \;\delta_i^j \; 
< P_F >(\eta) \; ,
$$
since we consider homogeneous and isotropic quantum states and density matrices.

More explicitly, the energy density in conformal time takes the form
\be\label{hf1}
 <{\cal H}_F>(\eta) \; = \;  \; <\Psi({\vec x},\eta)^{\dagger}\; H_F \;   \Psi({\vec x},\eta)>  \quad ,
\ee
where the fermion hamiltonian $ H_F $ is defined by
\be\label{hf2}
a(\eta) \; \gamma_0 \; H_F \;=\; -\;i {\vec \gamma}\cdot {\vec \nabla}\; + \;m \;  a(\eta) \;= \;
{\vec \gamma} \cdot {\vec p}\;+ \;m \;  a(\eta) \; .
\ee
An analogous expression can be written for the pressure,
\be\label{pf}
<P_F>(\eta) \;= \;\frac1{3 \;  a(\eta)} \; <{\bar \Psi} \; {\vec \gamma}\cdot {\vec p} \; \Psi>(\eta)\; .
\ee
Here too, it is convenient to consider the conformal energy and pressure,
\be\label{epconf}
{\varepsilon}_F(\eta) \;\equiv \; a^4(\eta) \; <{\cal H}_F>(\eta) \quad ,  \quad 
 p_F(\eta) \;\equiv \; a^4(\eta) \; < P_F >(\eta) \; .
\ee
We find the trace of the energy-momentum tensor from Eqs.(\ref{hf2}), (\ref{pf}) and (\ref{epconf}),
\bea\label{vir}
&& \varepsilon_F(\eta) \;- \;3 \; p_F(\eta) \;= \; m \; a(\eta) \; \Sigma_F(\eta) \; , \qquad  {\rm or} 
 \\ \cr 
&& <{\cal H}_F>(\eta) \;- \;3 \; <P_F>(\eta) \;= \;m \; <{\bar \Psi} \Psi>(\eta)  \; . \nonumber
\eea
This is the expression of the virial theorem in the present context and
\be\label{dfsf}
\Sigma_F(\eta)\; \equiv \; a^3(\eta) \; <{\bar \Psi} \Psi>(\eta) \; .
\ee
The above  expressions for the energy density and pressure obey
the usual continuity equation in cosmic time
\be\label{cont}
\frac{d}{dt} <{\cal H}_F> \;  + \;  3 \, H(\eta) \; \left( <{\cal H}_F> + <P_F> \right) = 0 \; ,
\ee
In conformal time by using Eqs.(\ref{vir})-(\ref{dfsf}) the continuity equation 
(\ref{cont}) becomes, 
\be\label{ceta}
\frac{d \varepsilon_F}{d\eta} \;= \;m \; \frac{d a(\eta)}{d\eta} \; \Sigma_F(\eta) \; .
\ee
We thus see from Eqs.(\ref{vir}) and (\ref{ceta}) that there is only one independent quantity among 
$ \varepsilon_F(\eta), \;  {\mathcal P}_F(\eta) $ and $ \Sigma_F(\eta) $.

\section{The Cosmological Quantum Vacuum.}\label{coqv}

There are two widely separate scales in the field evolution in the cosmological space-time.
The fast scale is the microscopic quantum evolution scale, typically $ \sim 1/M \sim 1/m $.
The slow scale is the Hubble scale $ 1/H $ of the universe expansion.

\medskip

When $ M \sim m \gg H $ we can consider that the scale factor is practically constant.
Therefore, in conformal time the quantum field evolution is like the evolution in Minkowski 
space-time with a mass $ M \; a(\eta) $ or $ m \; a(\eta) $ for bosons or fermions, respectively 
[see Eqs.(\ref{massdelta}) and (\ref{pamd})].

\medskip

The scalar and fermion densities follow as equal point limits of
the scalar and fermion two point functions. That is,  we consider the scale factor $ a $ as a constant
and obtain for the scalar two point function 
\bea\label{dosps}
G_S({\vec x}-{\vec x}', \eta-\eta',M \, a) &\equiv& 
<T \; \varphi({\vec x},\eta) \; \varphi({\vec x}',\eta')> \; = \frac1{a^2} \; 
\int \frac{d^4 k }{(2 \, \pi)^4} \; e^{-ik \cdot (x-y)} \frac{i}{k^2 -a^2 \; M^2 +i \, 0} = \cr \cr
&=&\frac1{(2 \, \pi)^2} \;  \frac{M}{z\; a} \; K_1(M \; a \; z)  \quad  , \quad 
z \equiv \sqrt{({\vec x}-{\vec x}')^2 - (\eta-\eta')^2} \; , \label{defz}
\eea
where $ K_1(x) $ is a modified Bessel function.

\medskip

Eq.(\ref{dosps}) is the zeroth order adiabatic approximation.
It differs from the exact two point function by quantities of the order $ {\cal O}(a'(\eta)),
\;  {\cal O}(a''(\eta)) $ etc. 

\medskip

We find from eq.(\ref{dosps}) in the coincidence point limit :
\be\label{p2ce}
 G_S({\vec x}-{\vec x}', \eta-\eta',M \, a) \buildrel{z \to 0}\over= \frac1{(2 \, \pi)^2} \,
\left\{\,\frac1{z^2 \; a^2} + \frac12 \, M^2 \,\left[\,\log\left(M \; a \; z \right) 
+  {\cal C}- \ln 2 -\frac12 \,\right] \right\} [\,1 +  {\cal O}(M^2 \, z^2 )\,] .
\ee
where $ {\cal C} = 0.57721566\ldots $ is the Euler-Mascheroni constant.
Eqs.(\ref{dosps})-(\ref{p2ce}) display the two point functions for the zero temperature vacuum.
The effect of a non-zero temperature on the two point function is negligible for
$ a \gg 1 $ as we show below [Eq.(\ref{Tcorr})].

\medskip

The fermion two point function takes the form 
\be \label{dospf}
<T \; \Psi({\vec x},\eta)_{\alpha} \; {\bar \Psi}({\vec x}',\eta')_{\beta}> \;\; = \;\; \frac1{a^3} \; 
\int \frac{d^4 k }{(2 \, \pi)^4} \; e^{-ik \cdot (x-y)} \;\; 
\frac{i\,({\not\!{k}}+ a \; m)_{\alpha \; \beta} }{k^2 -a^2 \; m^2 +i \, 0} \; ,
\ee
and hence,
\be \label{confe}
 G_F({\vec x}-{\vec x}', \eta-\eta',m \, a) \equiv \; 
<T \; {\bar \Psi}({\vec x},\eta) \; \Psi({\vec x}',\eta')>  \; = -\; 4 \, m \; 
G_S({\vec x}-{\vec x}',\, \eta-\eta',\,m\, a), \qquad  {\rm Dirac ~ Fermions} \; .
\ee
The minus sign in front arose from the anticommutation of the fermion fields going from 
Eq.(\ref{dospf}) to  Eq.(\ref{confe}). Here we used Eq.(\ref{dosps}) and
\be \label{tra}
{\rm Tr} \; {\not\!{k}} \;=\; 0 \quad , \quad {\rm Tr} \; 1\; = \;4 \; .
\ee
That is, the factor $ 4 = 2 \times 2$ in Eqs.(\ref{confe}) and (\ref{tra}) 
comes from the fermion and antifermion 
contributions times the number of helicities of a Dirac fermion.
Hence, this factor $ 4 $ becomes a factor $ 2 $ for Majorana fermions:
\be \label{confeM}
 G_F\,(\,{\vec x}-{\vec x}'\,,\, \eta-\eta'\,,\,m \, a\,) \;\equiv  \;\; 
<{\bar \Psi}({\vec x},\eta) \; \Psi({\vec x}',\eta')> \;\;  = \;-\, 2 \, m \; 
G_S\,({\vec x}-{\vec x}'\,,\, \eta-\eta',\,m\, a\,) \; .
\ee
We find in the coincidence point limit  up to corrections $ [\,1 +  {\cal O}(m^2 \; z^2 )\,] $:\;\;

\bea \label{p2cf}
&&G_F({\vec x}-{\vec x}', \eta-\eta',m \, a)\buildrel{z \to 0}\over= - \, 
\frac{2 \; \mathcal{N} \; m}{(2 \, \pi)^2} \;
\left\{\frac1{z^2 \; a^2}\,+\, \frac12 \; m^2 \,  \left[\,\log\left(m \; a \; z \right) 
+  {\cal C} - \ln 2 -\frac12 \,\right] \right\} 
\eea \;\;

Here, $ \mathcal{N} = 1 $ for Majorana fermions and  $ \mathcal{N} = 2 $ for Dirac fermions.

\bigskip

In order to define the vacuum densities as the coincidence limits,
$$
<\varphi^2>(\eta) \; \equiv \;  \;  <\varphi^2({\vec x},\eta)> 
\quad , \quad <{\bar \Psi} \Psi>(\eta) \; \equiv \; \; <{\bar \Psi}({\vec x},\eta) \; \Psi({\vec x},\eta)>
$$
we have to subtract the singularities at $ z = 0 $ in Eqs.(\ref{p2ce}) and (\ref{p2cf}).
Subtracting the singularities leaves a finite $z$ independent piece.
Requiring that the vacuum densities vanish in Minkowski space-time ($ a = 1 $) we obtain,
\be\label{denss}
<\varphi^2>(\eta) = \frac{M^2}{2 \, (2 \, \pi)^2} \; \left[\,\log a + b_S \; f_S(a) \, \right]
\quad , \quad <{\bar \Psi} \Psi>(\eta)  = - \frac{\mathcal{N} \; m^3}{(2 \, \pi)^2} \; \left[\, \log a
+ b_F \; f_F(a) \, \right] \; 
\ee
The functions $ f_S(a) $ and  $ f_F(a) $ are finite and vanish for Minkowski space-time,
$$
 f_S(1) \;= \;0 \quad , \quad f_F(1) \;= \;0 \; .
$$
We compute the terms  $ b_S \; f_S(a) $ and  $ b_F \; f_F(a) $
with the result:
$$
 f_S(\infty) \buildrel{ a(\eta) \;\gg \; a_{\rm dcs},\;  a_{\rm dcf}}\over= \;1 \;+ \;{\cal O}\left(\frac1{a^2}\right),
\quad  \quad f_F(\infty) \buildrel{ a(\eta)\; \gg \;a_{\rm dcs},\;  a_{\rm dcf}}\over= \;1 \;+ \;{\cal O}\left(\frac1{a^2}\right)
\; .
$$
When one performs an infinite subtraction at $ z = 0 $, an additional finite 
subtraction can always be done. We recognize that the additional terms containing $ b_S $ and  $ b_F $
can be absorbed in a finite multiplicative renormalization of the scale factor.
That is, introducing $ b_S $ and  $ b_F $ amounts to a scale transformation.
We compute the coefficients $ b_S $ and  $ b_F $ in terms
of the subtraction scale in momentum space $ (x \; M) $ for scalars and $ (x \; m) $ 
for fermions, with the result
$$
b_S(x)\; = \;b_F(x)\; =\; -\,\frac12 \;-\; \log x \;- \;\log a_{\rm dc} \; .
$$
where $ a_{\rm dc} $ stands for the scale factor at decoupling time (initial time).
In summary, we have for the late time regime,
\bea\label{condasi}
&&<\varphi^2>(\eta) \buildrel{ a(\eta)\; \gg \;a_{\rm dcs},\;  a_{\rm dcf}}\over= \frac{M^2}{2 \, (2 \, \pi)^2} \; 
\left[\,\log a(\eta) \,+ \,b_S \,\right] \; = \;\frac{M^2}{2 \, (2 \, \pi)^2} \; 
\left[\,\log \frac{a(\eta)}{x \; a_{\rm dcs} } - \frac12 \,\right],  \cr \cr 
&& <{\bar \Psi} \Psi>(\eta)  \buildrel{ a(\eta)\; \gg \;a_{\rm dcs},\;  a_{\rm dcf}}\over= 
- \frac{\mathcal{N} \; m^3}{(2 \, \pi)^2} \; \left[\,\log a(\eta)\, + \,b_F \, \right]\; = \;
- \frac{\mathcal{N} \; m^3}{(2 \, \pi)^2} \; 
\left[\,\log \frac{a(\eta)}{x\; a_{\rm dcf}} - \frac12 \,\right] \; .
\eea
where $ a_{\rm dcs} $ and $ a_{\rm dcf} $  stand for the scale factor at the decoupling times 
(initial times) for the scalar and the fermion, respectively.

The two-point functions Eqs. (\ref{dosps}) and (\ref{confe}) correspond to the zero temperature case.
The singular pieces for $ z \to 0 $ are temperature independent. We can disregard
the temperature dependent contributions to the two point functions since for large $ a $ they decrease as
\be\label{Tcorr}
  \sqrt{M} \; \left( \frac{T}{2 \, \pi \; a}\right)^{\frac32} \; e^{- \frac{M \; a}{T}} \,\to \,0 
\quad , \quad  a \gg 1 \; .
\ee
The scalar and fermion densities $ <\varphi^2>(\eta) $ and $ <{\bar \Psi} \Psi>(\eta) $
can be also computed as momentum integrals over the mode functions 
$ \phi_k(\eta) $ and $ f_k(\eta) $. 
In addition, the subdominant 
terms in $ 1/ a^2(\eta)\;, \; {\dot a}(\eta)/ a^2(\eta)\;, \ldots $ etc.
 can be obtained.
\bigskip

The equal points behaviour of the two point function Eqs.(\ref{p2ce}) and (\ref{p2cf}) 
is generic for any curved space-time when expressed as a function of 
the geodesic (squared) distance $ \sigma \equiv z^2 \; a^2 $ between the two points.
That is, the short distance behaviour is uniquely and universally determined by the local 
space-time geometry. It must be noticed that the divergences and finite pieces at $ \sigma = 0 $ 
are of the {\bf same type} as in Minkowski space-time. This is the so called Hadamard expansion for 
$ \sigma \to 0 $ and is equivalent to the adiabatic expansion. The coefficients of the divergent 
and finite parts are called Hadamard coefficients and they are known for generic space-times.

\section{Vacuum Energy Density and Pressure for Late Times}

The total energy density $ \varepsilon(\eta) $ and pressure $ {\mathcal P}(\eta) $: 
\be
 <{\cal H}>(\eta)\; = \;\; <{\cal H}_S>(\eta) \;\, +  \;<{\cal H}_F>(\eta) 
\ee
\be
 <P>(\eta) \; = \;\;   <P_S>(\eta) \, \;  + \; <P_F>(\eta),  
\ee
can be computed in the late time regime using the virial theorem Eqs. (\ref{virB}) and (\ref{vir}), 
the continuity equation Eqs. (\ref{cetaB}) and (\ref{ceta}) and the late time behaviour of 
the densities, Eq.(\ref{condasi}).

\bigskip

We obtain after calculation for the energy density and pressure,
\be\label{hp}
<{\cal H}>(\eta) \buildrel{ a(\eta) \;\gg\; a_{\rm dcs},\;  a_{\rm dcf}}\over = 
\frac{M^4}{2 \, (4 \, \pi)^2}\left[\log a(\eta) + b_S 
 -\frac14 \;\right] - 
 \frac{m^4}{(4 \, \pi)^2}\; \mathcal{N} \, \left[ \log a(\eta) + b_F  -\frac14 \right]
\ee
\be
 <P>(\eta)  \buildrel{ a(\eta)\; \gg \;a_{\rm dcs},\;  a_{\rm dcf}}\over= 
-\frac{M^4}{2 \, (4 \, \pi)^2}\;\left[ \log a(\eta)+ b_S +\frac1{12}\right]
+\frac{m^4}{(4 \, \pi)^2}\; \mathcal{N} \,\left[\log a(\eta) + b_F +\frac1{12}\right]   
\ee
The decoupling (initial) times for the evolution of scalars and fermions can be different
to each other. We have absorbed in  $ b_S $ and  $ b_F $ the corresponding initial
values of the scale factor for scalars and fermions, respectively.

The positivity of the energy density impose the condition,
$$
M^4\; > \;2 \; \mathcal{N} \; m^4 \; .
$$
Notice that,
$$
<P>(\eta) \;  + \; <{\cal H}>(\eta) \buildrel{ a(\eta)\; \gg \; a_{\rm dcs},  \;a_{\rm dcf}}\over= 
- \;\frac1{6 \, (4 \, \pi)^2}\;\left[\, M^4\; - \;2 \; \mathcal{N} \; m^4\,\right] ,
$$
is time independent and independent of the finite subtraction coefficients $ b_S $ and  $ b_F $
as well.

From Eq.(\ref{hp}) we obtain for the equation of state,

\be\label{asieces1}
w(\eta) \;\equiv\; \frac{<P>(\eta)}{ <{\cal H}>(\eta)} 
\buildrel{ a(\eta) \;\gg \;a_{\rm dcs},\;  a_{\rm dcf}}\over=  -\;1 \;- \;\frac13 \, \left[\;{\displaystyle \log a(\eta)\;-\;\frac14
\;+\;\frac{ b_S - {(\;2 \; \mathcal{N} \; m^4}/{M^4}\;) \; b_F}{1  - {(\;2 \; \mathcal{N} \; m^4} / {M^4}\;) }}\;\right]^{-1} 
\ee\;\;

That is, we find $ w(\eta) < - 1 $ with  $ w(\eta) $ asymptotically reaching the value
$ - 1 $ from below. 

\medskip

It is convenient to express the scale factor in terms of the redshift as
\be\label{az}
a(\eta) \; \exp{(b_S)} \;= \;\frac{1 \;+ \;z_S}{1 \;+ \;z} \qquad,  \qquad 
a(\eta) \; \exp{(b_F)}\; = \;\frac{1 \;+ \;z_F}{1 \;+ \;z} \; ,
\ee
where $ z_S $ ($ z_F $) is the redshift when the evolution of the scalar (fermion) become the 
one of a free field in the cosmological space-time. In terms of  $ z_S $ and $ z_F $ Eqs.(\ref{hp}) read, 
\be\label{hpz}
\begin{split}
[\; 2\,(4 \,\pi)^2 \;]\, <{\cal H}>(z) \; = \; M^4\,  \log\;(1 + z_S)\;
-\; 2 \,\mathcal{N}  m^4 \, \log\; (1 + z_F)\; - \\ -\;( M^4 \,-\, 2 \, \mathcal{N} m^4) \left[ \;\log\,(1 + z) \;+ \;\frac14 \;\right]\,.  
\end{split}
\ee
\be\label{hpz2} 
\begin{split}
[\;-\; 2\; (4 \; \pi)^2\;]\; <P>(z) \; = \; \; M^4 \; \log \; (\;1 + z_S\;)\; 
- \; 2 \; \mathcal{N} \; m^4 \; \log\;(\; 1 \;+ \; z_F\;) \; - \\
 - \;(\; M^4 \;- \; 2 \; \mathcal{N} \, m^4\;)\,\left[\,\log \;(\;1 \;+\; z\;) \;- \;\frac1{12} \,\right]\,
\end{split}
\ee
The equation of state (\ref{asieces1}) as a function of $ z $ takes the form:
\begin{multline}\label{wm1z}
w(z)\, + \,1 \;= \; - \frac13 \, \left( \; 1  - \frac{2 \, \mathcal{N} \, m^4} {M^4} \;\right) \times\; \\
\times \;\left\{\, \log\; (1 + z_S) - 
\left(\frac{2 \, \mathcal{N} \, m^4} {M^4}\right) \log\;(1 + z_F) - \left(1  -  \frac{2 \, \mathcal{N} \, m^4} {M^4}\right)\left[\, \log\;(1 + z) + \frac14 \,\right ] \right\}^{-1}  
\end{multline}

The equation of state and the energy density become today:
\be\label{ecshoy}
w(z = 0) + 1 \;= \; - \frac13 \, \left(1  - \frac{2 \; \mathcal{N} \; m^4} {M^4}\right) \left\{\,\log\,(1 + z_S) -\frac14-
\left(\frac{2 \; \mathcal{N} \; m^4} {M^4}\right) \left[\,\log\,(1 + z_F) -\frac14 \,\right] \right\}^{-1}
\ee
\be \label{dehoy2}
 <{\cal H} > (z = 0) \;= \;\frac1{2 \, (4 \, \pi)^2}\left\{ M^4 \left[\, \log\,(1 + z_S) -\frac14 \,\right] 
-2 \; \mathcal{N} \; m^4 \left[\,  \log\,(1 + z_F)  -\frac14 \,\right] \right\} 
\ee
The energy density at late times $ \eta $ after decoupling and the energy density today are related by 

\be 
<{\cal H}>(\eta) \buildrel{ a(\eta)\; \gg \;a_{\rm dcs},\;  a_{\rm dcf}}\over =  <{\cal H}>(z = 0) \; + \;
\left(\frac{M^4 - 2 \, \mathcal{N} \; m^4}{2 \, (4 \, \pi)^2}\right)\; \log \left(\frac{a(\eta)}{a_0}\right) 
\ee 

where we used Eqs.(\ref{hp}) and (\ref{ecshoy}) and $ a_0 $ stands for the scale factor today. 

We identify the vacuum energy density today 
$ <{\cal H}>(z = 0) $ with the observed dark energy $ \rho_{\Lambda} $.
We can then write,
\be\label{rhoeta}
<{\cal H}>(\eta)\; = \; \rho_{\Lambda} \left[\, 1 + \beta_{\mathcal{N}} \;  \log \frac{a(\eta)}{a_0}\, \right] 
\ee
where
\be\label{defbeta}
\beta_{\mathcal{N}}\,\equiv \, \left(1  - \frac{2 \; \mathcal{N} \; m^4}{M^4} \right) \left\{\,\log\,(1 + z_S)\, - \,\frac14
\,- \,\left(\frac{2 \; \mathcal{N} \; m^4}{M^4}\right) \left[\, \log\,(1 + z_F) \,- \,\frac14 \,\right] \,\right\}^{-1}
\ee
That is, the vacuum energy density at late times after decoupling {\bf grows} as the logarithm
of the scale factor. Moreover, the equation of state approaches $ -1 $ from below as:

$$
w(\eta) + 1 \buildrel{ a(\eta) \;\gg\; a_{\rm dcs}, \; a_{\rm dcf}}\over= - 
\left(\frac{M^4 - 2 \, \mathcal{N} \; m^4} {6 \, (4 \, \pi)^2 \;\rho_{\Lambda}}\right)
\left[\, 1 \,+ \,\beta_{\mathcal{N}} \;  \log \frac{a(\eta)}{a_0}\, \right]^{-1} .\;\;
$$

The previous equations in this subsection generalize when there are several scalar and fermion
fields just summing over their respective contributions. Let us consider the case
of several scalars and fermions: {\bf This case} is relevant to study whether the three neutrino mass
eigenstates can contribute to the dark energy. Eqs.(\ref{ecshoy}) become for $ z_S , \; z_F \gg 1 $:

\bea\label{3nu}
&& w(z \,=\,0)\; + \;1 \;= \; - \;\frac{\sum_j M_j^4\;- \;2 \; \mathcal{N} \; \sum_i  m_i^4}{6 \, (4 \, \pi)^2 \; \rho_{\Lambda} }
\; , \\ \cr
&& \rho_{\Lambda} \;= \;\frac1{2 \, (4 \, \pi)^2}\left[ \left( \log z_S -\frac14 \right) \sum_j M_j^4
-2 \; \mathcal{N}\left( \log z_F  -\frac14 \right) \sum_i  m_i^4 \right] \;  \label{3nu2}
\eea

where $ j $ and $ i $ label the species of scalars and fermions, respectively. 

It is convenient to eliminate the 
sum of scalar masses $ \sum_j M_j^4 $ between eqs.(\ref{3nu}) and (\ref{3nu2}) with the result,
\be\label{3nu3}
 w(z\, =\,0)\; + \;1 \;= \; \frac1{ (\log z_S -\frac14)} \left[\; -\frac13 \; + \; \frac{\mathcal{N}}{3 \, (4 \, \pi)^2}
\; \frac{\sum_i  m_i^4}{\rho_{\Lambda}} \; \log \frac{z_S}{z_F} \; \right] \; 
\ee \;

We see in Eq.(\ref{3nu3})  that the scalar contributes to the equation of state today by the negative term
$ -1/[\,3 \, (\,\log z_S -\frac14\,)\,] $ while the fermions give for $ z_S > z_F $
a {\it positive} contribution proportional to the sum of the fourth power of their masses.

\section{The Quantum Nature of the Cosmological Vacuum.}

Local observables as $ <\varphi^2>, \;   <{\bar \Psi} \Psi> $, the energy density and the pressure
involve the product of field operators at equal points. This is identical to one-loop tadpole 
Feynman diagrams. Logarithmic dependence on the scale of the momenta is typical in 
one-loop renormalized Feynman diagrams \cite{IZ}. Here, we analogously find
a logarithm of the scale factor in Eqs.(\ref{condasi}) and (\ref{hp}) 
through the same mechanisms at work in renormalized quantum field theory.
Hence, the dark energy follows here as a truly {\it quantum field vacuum effect}.
We stress quantum {\it field} effect and not just quantum effect because the infinite number of 
filled momentum modes in the {\it vacuum} as well as the
subtraction of UV divergences play a crucial role in the vacuum late time behaviour.
Here the quantum fields are not coupled neither self-coupled but they interact
with the expanding space-time geometry.

\bigskip

Notice that these results Eqs.(\ref{condasi}), (\ref{hp}) and (\ref{asieces1})
 are valid for {\bf any} expanding universe.
They do not depend on the specific time dependence of the scale factor
$ a(\eta) $ provided it grows with $ \eta $. 

\bigskip

The quantum nature of the  vacuum cosmological effects in the
physical observables here are manifest from Eqs.(\ref{condasi}) and (\ref{hp}),
\bea \label{densn}
<\varphi^2>(\eta)&&\sim  \quad M^2 \; \log  a(\eta) \; =\;
\frac{M^2 \; c^2}{ \hbar} \; \log  a(\eta)\; = \; \frac{M \; c}{\lambda_C}
\; \log  a(\eta) \; , \cr \cr 
<{\cal H}>(\eta) &&\sim  \quad M^4 \; \log  a(\eta)  
\;= \;M \; c^2 \left( \frac{M \; c}{\hbar}\right)^3  \; \log  a(\eta) 
\;= \; \frac{M \; c^2}{\lambda_C^3}  \; \log  a(\eta) \; 
\eea
These quantities are of quantum nature since they depend on $ \hbar $.
There is no `classical contribution' to the vacuum energy. Eq.(\ref{densn}) just means
that the  scale of the dark energy density is of one scalar rest mass per 
a volume equal to the cube of the Compton wavelength $ \lambda_C $
for the scalar particle.
Notice that $ \lambda_C = [\hbar /{(M \; c)}] \simeq 0.05$ mm is almost a macroscopic length
while the mass of the scalar particle $ M \sim 4 $ meV $ = 7.1 \, 10^{-36}$ g is extremely small
(see below for the value of $ M $). 

\section{Dark Energy from the Cosmological Quantum Vacuum.}

Let us recall the current value for the dark energy density 
\be\label{dato}
\rho_{\Lambda} \;= \; \Omega_{\Lambda} \; \rho_c \;= \;3.28  \times  10^{-11} \; ({\rm eV})^4
\;= \;(2.39 \;  {\rm meV})^4 \; ,
\ee
corresponding to 
\be\label{h}
h\; = \;0.73 ,\qquad \Omega_{\Lambda} \;= \;0.76 , \quad \;\text{and  where}\;\; 1 \; {\rm meV}\;=\; 10^{-3}\; {\rm eV}.
\ee
 
We take these values because they do correspond to {\it direct},  {\it model independent} and {\it late} universe observations,  Refs \cite{descu}, \cite{schmidt}, \cite{scp},  \cite{mas}, \cite{DES},  \cite{Riess2020}, \cite{Riess2021}, \cite{Riess2022},
  and accordingly this paper deals directly with dark energy in the late universe; moreover, dark energy was discovered with such direct and model independent measurements in the late universe, Refs \cite{descu}, \cite{schmidt}, \cite{scp}, \cite{mas}.   Other determinations of $h$ (eg. Ref \cite{Planck2020} Table 2, page 16) yield values $h = 0.68 $, $\Omega_{\Lambda} = 0.69 $. However, these are 
 indirect, model dependent and early universe determinations of $h$ and $\Omega_{\Lambda}$. The difference between the determinations of $h$ in the late and in the early universe is an  important problem by its own, eg. Ref. \cite{Riess2022}, although we do not treat this problem here.

\bigskip

Bosons give a positive contribution to the dark energy through the cosmological quantum vacuum 
while fermions give a negative contribution. Therefore, the boson contribution {\it must} dominate.

As discussed in Sec. \ref{estneu}, the lightest neutrino certainly contributes to the  
cosmological quantum vacuum unless it dissipates. Definitely, 
a boson contribution {\it is needed}. The photon and graviton contributions are irrelevant
since their masses are most probably zero and at most $
m_{\;\gamma} \,< \,6 \times 10^{-17}  \;  {\rm eV},  \quad 
m_{\;\rm graviton} \,< \,4.7 \times 10^{-23}  \;  {\rm eV} $\;  \cite{Abbott2019}.

Massless particles contribute to the energy-momentum tensor through the trace anomaly \cite{BD}, \cite{anom}.
This contribution is of the order of $ H_0^4 $ where $ H_0 $ is the Hubble parameter today:
\be\label{hubhoy}
 H_0 \;= \;1.558 \,\times \,10^{-33} \;  {\rm eV} \; ,
\ee
As a consequence, the massless particles contribution to the energy-momentum tensor is exceedingly small to explain the observed value of the dark energy.

\begin{itemize}
\item{A scalar particle can produce the dark energy today Eq.(\ref{dato}) through its quantum
cosmological vacuum provided:}
\item{Its  mass is of the order of $ 1 $ meV and it is very weakly coupled.}
\item{Its lifetime is of the order of the age of the universe.}
\end{itemize}

Spontaneous symmetry breaking of continuous symmetries produces massless
scalars as Goldstone bosons. If, in addition, this continuous symmetry is
slightly violated the Goldstone boson acquires a small mass.
This is the natural mechanism that generates light scalars and several
particles have been proposed on these grounds in the past.
The axion is certainly the one that caught more attention in the literature.
Other proposed particles are the familons and the majorons \cite{majo}, \cite{rneu}. 

The (invisible) axion \cite{invi} (if it exists) is hence a candidate to be the source of dark energy.

\medskip

Axions were proposed to solve the strong CP problem in QCD \cite{axi0}. Axions
acquire a mass after the breaking of the Peccei-Quinn (PQ) symmetry when the temperature
of the universe was at the PQ symmetry breaking scale 
$ \sim f_a $ \cite{axi}. All axion couplings are inversely proportional to $ f_a $ and
the axion mass is given by 
\be
M_a \; \simeq \; 6 \; \times \; \left(\frac{10^9 \;  {\rm GeV}}{f_a} \right) {\rm meV} \; .
\ee
The following range (`axion window') is currently acceptable for the axion mass \cite{axiM}, \cite{axiM2}:
\be\label{cotma}
10^{-3} \;  {\rm meV} \; \lesssim \; M_a \; \lesssim \; 10 \;  {\rm meV} \; 
\ee
Therefore, this pseudoscalar particle has extremely weak coupling to gluons and quarks 
and hence it contributes to the cosmological quantum vacuum. For example, the
axion-photon-photon coupling is given by 
\be
g_{a \; \gamma \; \gamma} \;\sim \; \frac{10^{-10}}{\rm GeV } \; \left(\frac{M_a}{1 \; {\rm meV}}\right) \; 
\ee

As a consequence, the axion lifetime to decay into photons is much longer than the age of the universe.
Dissipation of the energy in the  cosmological quantum axion vacuum
takes longer than the age of the universe too.
\begin{itemize}
\item{An axion with mass $ \sim 1 $ meV and hence $ f_a \sim 10^9 $ GeV 
decoupled from the plasma at a scale of energies
$ \sim 2 \times 10^5$ GeV , that is at redshift 
$ z_S \sim 2.2 \times 10^{18} $. The temperatures of the axions and neutrinos 
today are lower than that of photons today,
\be \label{tnuhoy}
T_{\nu\; \rm today} \;= \;\left( \frac4{11} \right)^{\frac13} \, T_{\rm CMB \; today}\; = \,
0.1676 \; {\rm meV} \quad , \quad T_{a \, \rm today} \;= \; 0.078 \; {\rm meV} \; .
\ee
Because the axion lifetime is of the order or larger than the age of the universe,
no specific properties of the axion play a role in the dark energy except for its mass
and decoupling redshift. However, the dark energy depends on the decoupling redshift 
rather weakly because it is through its logarithm [see Eq. (\ref{hpz})].}
\item{Neutrinos in the universe are believed to be effectively free particles when the temperature
of the universe is below $ \sim 1$ MeV. That is, neutrinos decouple at
a redshift $ z_F \sim 0.6 \times 10^{10}$. 
Before such time, electrons and neutrinos interact, keeping them in thermal equilibrium.}
\item{Therefore, we can treat the axion  with mass $ \sim 1 $ meV and the lightest neutrino as free particles in the universe 
for redshifts $ z < z_S \sim 2.2 \times 10^{18} $ and $ z <  z_F \sim 0.6 \times 10^{10}$, respectively.}
\end{itemize}

\section{Neutrino Mass Eigenstates}\label{estneu}

As is known, the two heavier neutrino mass eigenstates $ \nu_2 $ and $ \nu_3 $ 
with masses $ m_2 $ and $ m_3 $, respectively, annihilate with their
respective anti-neutrinos yielding the lightest neutrino eigenstate $ \nu_1 $ 
and its antiparticle through weak
interactions. However, this process is too slow for nonrelativistic neutrinos
even compared with the age of the universe. Their decay rates can be estimated to be
$$
\Gamma_2 \;\sim \;G_F^2 \; m_2^5 \; \sim \; \frac1{1.5 \;\times\; 10^{33} \; {\rm yr}} \quad , \quad
\Gamma_3 \;\sim \;G_F^2 \; m_3^5 \;\sim \; \frac1{5 \; \times\; 10^{29} \; {\rm yr}}  \; .
$$
where $ G_F = 1.166 \;\times \; 10^{-23} \; ({\rm eV})^{-2} $ stands for the Fermi coupling.

Neutrinos with masses $ m_2 \sim 0.01$ eV or  $ m_3 \sim 0.05$ eV will produce
through their cosmological quantum vacuum today a large negative contribution to the dark energy. 

Therefore, the heavier neutrinos ($ \nu_2 $ and $ \nu_3 $) must
annihilate with their respective anti-neutrinos into the lightest neutrino  
$ \nu_1 $ through a mechanism such that  
\be \label{conG}
\Gamma_3 \;\;\gtrsim \;\;\Gamma_2 \;\;\gtrsim \; \; {\rm (age \; of \; the\;  universe)}^{-1}  \; .
\ee
Our results for the dark energy are {\it independent} of the details of the decay mechanism.
All what counts is that the decay rates of the heavier neutrinos obey Eq.(\ref{conG}).

\medskip

As a minimal assumption, let us consider the following effective
couplings between the neutrinos,
\be\label{lanuf}
\frac1{M'\,^2} \; {\bar \Psi}_2  \;  \Psi_2  \;  {\bar \Psi}_1  \;  \Psi_1
\quad \;, \quad \;
\frac1{M'\,^2} \; {\bar \Psi}_3  \;  \Psi_3  \;  {\bar \Psi}_1  \;  \Psi_1 \; ,
\ee
where $ M' $ is a mass scale much larger than the neutrino masses.
We thus find,
$$
\Gamma_2 \,\sim \, \frac{(m_2)^5}{M'\;^4}  \quad \;, \quad \;
\Gamma_3 \, \sim \,\frac{ (m_3)^5}{M'\;^4}     \; .
$$
Imposing Eq.(\ref{conG}) yields,
\be\label{masfi}
 M'\;\lesssim \;1 \; {\rm MeV} \;  \; {\rm for~} m_2 \, = \, 0.01 \; {\rm eV} \quad
 {\rm and} \quad M' \;\lesssim \;10 \; {\rm MeV} \;  \; {\rm for~} m_3 \,= \,0.05 \; {\rm eV} \; 
 \ee
The first estimated bound ($ 1 $ MeV) applies for a direct hierarchy of neutrino masses 
($ m_3\sim  0.05 \; {\rm eV} > m_2 \sim  0.01 \; {\rm eV} > m_1 $) while
the second estimate  ($ 10 $ MeV) is for an inverse hierarchy of neutrino masses
($ m_3 \sim m_2 \sim  0.05 \; {\rm eV} > m_1 $).

\medskip

Effective couplings of the type  Eq.(\ref{lanuf}) can be obtained from different renormalizable
models.

Notice that the two heavier neutrinos decays contribute to the background of 
lighter neutrino particles but not to the neutrino quantum vacuum.

\bigskip

Lagrangians leading to effective couplings analogous to Eq.(\ref{lanuf}) have been
considered in the context of models to generate neutrino masses and to provide
light dark matter candidates \cite{matosc}. Moreover, mass ranges compatible
with Eq.(\ref{masfi}) have been obtained from various and independent considerations \cite{marke}, \cite{matosc}.
In case the  effective couplings  Eq.(\ref{lanuf}) arise from
Yukawa couplings of the neutrinos with a scalar particle of mass $ M' $, this scalar particle cannot be a dark matter candidate since it decays into neutrino-antineutrino pairs.

\bigskip

The lightest neutrino with mass $ m_1 $ can be self-coupled through the interaction
$$
\frac1{M''\,^2}   \;  \left({\bar \Psi}_1  \;  \Psi_1\right)^2
$$
Its decay rate,
$$
\Gamma_1 \;\sim \; \frac{(m_1)^5}{M''\,^4} 
$$
is of the order or larger than the age of the universe when
\be\label{autoa}
M'' \; \lesssim \; \left(\frac{m_1}{\rm meV}\right)^{\frac54} \; 50 \; {\rm keV} \; .
\ee
Hence, if Eq.(\ref{autoa}) is fulfilled, the energy in the 
neutrino vacuum dissipates into the lightest neutrinos $ \nu_1 $, thus 
contributing to the neutrino background.

\section{Light Particle Masses and the Dark Energy Density Today}\label{mM}

Let us consider the case where only one light scalar field and one
light fermion field contribute to the quantum vacuum energy.  That is,
a light scalar and the lightest neutrino. We obtain from Eq.(\ref{ecshoy}) 
for the mass of the scalar,
\be \label{Mgen}
M \;=\; \frac{2^{\frac54} \; \sqrt{\pi} \;  
\rho_{\Lambda}^{\frac14} }{\left(\log z_S\,- \,\frac14\right)^{ \! \frac14}}\,
\left[\; 1 \;+ \;\frac{\mathcal{N} \; m^4}{(4 \, \pi)^2 \; \rho_{\Lambda}} \; 
\left(\;\log z_F-\frac14 \; \right)\;\right]^{\frac14} \; 
\ee
where we identified the vacuum energy density today 
$ <{\cal H}>(0) $ with the observed dark energy $ \rho_{\Lambda} $.

We now obtain using the observed value of the dark energy Eq.(\ref{dato}) and
the decoupling redshift for the neutrino  $ z_F \sim 0.6 \times 10^{10}$,
\be\label{cota1}
M \;= \;\frac{10.1 \; {\rm meV}}{\left(\, \log z_S \,- \,\frac14 \right)^{ \! \frac14}}\,
\left[\; 1 \;+ \; \mathcal{N} \; \left(\; \frac{m}{3.90 \; {\rm meV}} \; \right)^4  \;\right]^{\frac14} 
\ee

If the lightest neutrino has a very small mass  $ m \ll 1 $ meV or if it
decays in the time scale of the age of the universe
[see Eq.(\ref{autoa})] so the neutrino vacuum dissipates, there is no neutrino 
contribution to the dark energy. In these cases Eq.(\ref{cota1}) gives for the mass of the scalar:
\be\label{sinnu}
M \;= \;\frac{10.1 \; {\rm meV}}{\left(\, \log z_S \;- \;\frac14 \,\right)^{ \! \frac14}}
\quad : \quad {\rm no ~ vacuum ~ neutrino ~ energy } \; .
\ee

\bigskip

Assuming the scalar field to be the axion we can use  the value $ z_S \sim 2.2 \times 10^{18} $ 
for the axion decoupling redshift and Eq.(\ref{cota1}) becomes,
\be\label{negma}
M(m) \;= \;3.96 \;  {\rm meV} \left[\; 1 \;+\; \mathcal{N} \; 
\left(\,\frac{m}{3.90 \; {\rm meV}}\,\right)^4 \;\right]^{\frac14} \; .
\ee
The values of the neutrino masses are not yet known, only their differences are experimentally
constrained. Both in the direct and inverse mass hierarchies the mass $ m $ of the 
lightest neutrino can be in the meV range (or even to be zero). 

According to Ref. \cite{masanu}, we have
$$
m \;= \;\frac13 \; m_2
$$
where $ m_2 $ is the mass of the middle neutrino. Combining this with the known neutrino
mass differences yields 
\be\label{mneu}
m \;= \;3.2 \;\pm \; 0.1  \; {\rm meV}
\ee
This value for the neutrino mass perfectly agrees in order of magnitude with the
see-saw prediction 
$$
 \frac{M^2_{\;\;\rm Fermi}}{M_{\;\rm GUT}}\; \simeq \; 6 \;\times \;10^{-3} \;  {\rm eV} \; ,
$$
for the typical values $ M_{\;\rm Fermi} \;= \;250$ GeV and $ M_{\;\rm GUT} \,= \, 10^{16}$ GeV,
of the Fermi and Grand Unified energy scales, respectively.

Eqs.(\ref{negma})-(\ref{mneu}) gives for the axion mass:
\be\label{majdir}
M\;( m = 3.2 \; {\rm meV},\;\mathcal{N} = 1 ) \;= \;4.35 \; {\rm meV} \quad ;  \quad
M\;( m = 3.2 \; {\rm meV},\;\mathcal{N} = 2 ) \;= \;4.66 \; {\rm meV} \; ,
\ee
for Majorana and Dirac neutrinos, respectively. 

If the lightest neutrino has a very small mass $ m_1 \ll 1 $ meV or if 
it decays in the time scale of the age of the universe 
[see Eq.(\ref{autoa})], eg there is no neutrino 
contribution to the dark energy, then the axion mass is given by 
\be\label{sinu}
M \;= \;3.96  \; {\rm meV} \quad : \quad {\rm no ~ vacuum ~ neutrino ~ energy } \; .
\ee
All the axion mass values Eq.(\ref{negma}) and Eqs.(\ref{majdir})-(\ref{sinu}) found here
describe the dark energy observed today Eq.(\ref{dato}).
The numerical values for the axion mass in Eqs.(\ref{majdir})-(\ref{sinu}) are
within the astrophysical bound of Eq.(\ref{cotma}).

\bigskip

We  compute the equation of state today from Eq.(\ref{ecshoy}) and 
display it in Table 1 in three relevant cases: 
(i) no neutrino contribution to the dark energy, 
(ii) a Majorana neutrino contribution,
(iii) a Dirac neutrino contribution. 
In all three cases the observed value  Eq.(\ref{dato}) of the dark energy is imposed.
For the last two cases we choose for the neutrino mass $ m = 3.2 $ meV and the scalar mass $M$
given by Eq.(\ref{majdir}), eg $4.35$ meV and $4.66$ meV respectively.

We see that $ w(0) $ is slightly below $ - 1 $ by an amount ranging
from $ (-1.5 \times 10^{-3} )$\; to \;$ (-8 \times 10^{-3}) $.

\bigskip

It can be noticed that the mass of the lightest neutrino [Eq.(\ref{mneu})] turns to be much higher
than today's neutrino temperature:
\be\label{ratmn}
\frac{m_{\;\rm Dirac}}{T_{\;\nu \;\rm today}} \;= \;19.6 \quad ,  
\quad \frac{m_{\;\rm Majorana}}{T_{\;\nu \;\rm today}} \;= \;23.3 \; ,
\ee
where we used Eq.(\ref{tnuhoy}). That is to say, the neutrinos forming the neutrino background
are today non-relativistic particles.

\bigskip

\begin{tabular}{|c|c|c|}\hline
  & & \\ Neutrino Type    & \;\;Scalar Mass\;\; & \;\;\;Equation of state today \;\;\;\\
 &  & \\ \hline & & \\
\;No ~ vacuum ~ neutrino ~ energy\;\; & $ M = 3.96 $ meV\; & $ w(0) + 1 = -.00794 $ \\ & & \\ \hline & &
\\ Majorana Neutrino  & \;\;$ M = 4.35 $ meV\;\; & $ w(0) + 1 = -.00473 $ \\  $ m = 3.2 $ meV & &
\\ \hline & & \\ Dirac  Neutrino  & \;\; $ M = 4.66 $ meV\;\; & $ w(0) + 1 = -.00156 $ \\  
$ m = 3.2 $ meV & &  \\ \hline
\end{tabular}

\bigskip

{\bf TABLE 1}. {\it The Equation of State Today   w(0) + 1 } computed from Eq.(\ref{ecshoy}) 
in three relevant cases which all describe the dark energy observed today [Eq.(\ref{dato})]: 
{\it (i)} No neutrino contribution to the dark energy;
{\it (ii)} A Majorana neutrino contribution with mass $ m = 3.2 $ meV;
{\it (iii)} A Dirac neutrino contribution with mass $ m = 3.2 $ meV. See the discussion in Sec. \ref{mM}.

\bigskip

Let us now analyze the possibility in which all three neutrino eigenstates contribute to the
dark energy. This contribution crucially depends on the values of their masses to the power
four through the dimensionless factor
$$
{\cal F}\; \equiv\; \frac1{3 \, (4 \, \pi)^2}
\; \frac{\sum\;_i  m_i^4}{\rho_{\Lambda}}
$$
as we see from Eqs.(\ref{3nu})-(\ref{3nu3}). 

For the normal hierarchy, we have 
$$
m_1 \;= \;3.2 \; {\rm meV} \quad , \quad m_2 \;= \;9.5 \; {\rm meV} \quad , \quad
m_3 \;= \;47 \; {\rm meV} \quad ,
$$
and for the inverted hierarchy: 
$$
m_1 \;= \;3.2 \; {\rm meV} \quad , \quad m_2 \;= \;47 \; {\rm meV} \quad , \quad
m_3 \;= \;48 \; {\rm meV} \quad .
$$
Thus, using Eqs.(\ref{dato}) the factor $ \cal F $ takes the values
$$
{\cal F}_{\rm normal}\; = \;315 \quad , \quad {\cal F}_{\rm inverted} \;= \;656 \; .
$$
Inserting these numbers in the equation of state today Eq.(\ref{3nu3}) yields
 values for $ w(0) $ in strong disagreement with the data unless we fine tune
$ z_S \simeq z_F $. Because  there is no reason to have such equality, we conclude that
the vacuum of the two heavier neutrinos must not contribute to the dark energy.
Their quantum vacuum must dissipate as discussed in Section \ref{estneu}.

\section{The Future Evolution of the Universe}

The future evolution of the universe follows by inserting the total energy
density in the Einstein-Friedmann equation 
$$
H^2(t) \;= \;\frac{8 \, \pi \; G}3 \; {\cal H}_T \; ,
$$
where we use cosmic time $ t , \; G $ is the gravitational constant and the total energy density
$ {\cal H}_T $ is the sum of the contributions from the dark energy, the matter and the radiation.

\medskip

We obtain using the dark energy expression Eq.(\ref{rhoeta}) the {\bf self-consistent}
Einstein-Friedmann evolution equation,
\be\label{fute}
H^2(t) \;= \;H_0^2 \,\left[\, \Omega_{\Lambda} \; \left(1 + \beta_{\mathcal{N}} \;
\log \frac{a(t)}{a_0}\right)\; +\; 
\Omega_{\rm matter} \; \frac{a^3_0}{a^3(t)} \;+\;
\Omega_{\rm rad} \; \frac{a^4_0}{a^4(t)} \,\right]
\ee
where $ a_0 \equiv a({\rm today}), \; \beta_{\mathcal{N}} $ is defined by Eq.(\ref{defbeta}),
$ \rho_{\Lambda} = \rho_{\rm crit} \; \Omega_{\Lambda} $ is given by Eq.(\ref{3nuI}), $H_0$ is the Hubble parameter today being: 
\be
\rho_{\rm crit}\; =\;
\frac{(3 \, H_0^2)} {(8 \, \pi \ G)} ,\qquad  H_0 \;=\; \frac{h} {[\;9.77813 \;  {\rm Gyr}\;]} ,\qquad
\Omega_{\Lambda}\; = \;0.76\; =\; (1 \,-\,\Omega_{\rm matter} \, -\,\Omega_{\rm rad}).
\ee

We get using the explicit values for $ M $ and $ m $ Eqs.(\ref{mneu})-(\ref{sinu}):
$$
\beta_0 \;=\; 0.0238:\quad {\rm No ~ vacuum ~ neutrino ~ energy };\quad  \quad 
\beta_1\; =\; 0.0347:\quad {\rm Majorana ~  neutrino} $$ $$
\beta_2 \;= \;0.0459: \quad {\rm Dirac ~   neutrino} \; .
$$

For $ a(t)\; \gtrsim \;a_0 $, the matter and radiation contributions can be neglected in Eq.(\ref{fute}) 
and we have,
$$
\left[\, \frac{d \; \log a(t)}{d \, t} \,\right]^2 \,\simeq \, H_0^2 \; \Omega_{\Lambda} \; 
\left[\, 1 + \beta_{\mathcal{N}} \; \log \frac{a(t)}{a_0} \,\right] ,
$$

\bigskip

This equation can be immediately integrated with the solution

\be\label{afute}
a(t)\;\buildrel{  H_0 \; t\; \gtrsim\; 1}\over\simeq a_0 \; 
\exp{[\,c_1 \;  H_0 \; t \; + \; c_2 \;  (H_0 \; t)^2\,]} \; ,
\ee
where 
\bea\label{bc}
&& c_1 \; = \;\sqrt{\Omega_{\Lambda}}\, = \,0.87 \quad , \quad
 c_2 \; = \; \frac14 \; \Omega_{\Lambda} \; \beta_{\mathcal{N}}\, = \, 0.19 \; \beta_{\mathcal{N}} \; ,\cr \cr
&& 0.00452\; < \; c_2 \;< \; 0.00872 \; .
\eea
The left and right ends of the interval in $ c_2 $ Eq.(\ref{bc})
correspond to the cases in which there is no neutrino contribution and to the lightest neutrino being
a Dirac fermion with mass $ m = 3.2 $ meV, respectively.

\medskip

We find that the Universe is presently reaching an asymptotic phase where it expands 
as indicated by Eq.(\ref{afute}).

\medskip

Eq.(\ref{afute}) shows that the expansion of the Universe 
in the future is {\it faster} than in the de Sitter Universe.

\medskip

Notice that the time scale of the accelerated expansion is huge $ \sim (1/H_0)  = 13.4$ Gyr.
The quadratic term dominates over the linear term in the exponent of  Eq.(\ref{afute})
by a time $ t \sim  (100/H_0)$ to $(200/H_0)$.

In this accelerated universe, Eq.(\ref{fute}) shows that the Hubble radius 
$( 1/H )$ {\it decreases} with time as 
$$ 
 \frac1{H} \;\sim \; \frac1{ H_0 \; \sqrt{\,\log a(t)}} \; .
$$

\section{Discussion}\label{discu}

The non-trivial energy and pressure that we have is an effect resulting of the expansion
of the space-time as it arises from the  $ \log a(\eta) $ factor in Eqs.(\ref{hp}). 
No dark energy appears in Minkowski space-time. Namely, the formation and growth of the vacuum density,
the vacuum energy density and pressure is an effect due to the presence
of quantum fields in an expanding cosmological space-time. 

\bigskip

Notice that the energy scale of the cosmological vacuum is given by the mass of the
particle when this mass is larger than the Hubble constant [see Eq.(\ref{hubhoy})].
For massless particles, the energy scale of the  cosmological vacuum
is given by the Hubble constant.

\bigskip

The axion evolution for $ z \geq 10^{18} $ as well as the  neutrino evolution for $ z \geq 10^{10} $
are beyond the scope of this article. Namely, the regime where the interaction of axions 
and neutrinos with the plasma particles cannot be neglected. We choose as initial state 
for both the axions and the neutrinos the vacuum thermal equilibrium state.
It must be remarked that the vacuum energy at late times is independent of the
initial temperature as shown by Eq.(\ref{Tcorr}).

\medskip

Before decoupling, particle interaction is non-negligeable and 
dissipation is important depleting the vacuum energy \cite{nosb}, \cite{nosf}.
Hence, the vacuum energy can only become significant after decoupling.
Therefore, it is a good approximation to just study
the free quantum field evolution in the cosmological space-time after decoupling.

The initial conditions  Eqs.(\ref{ciB}) and (\ref{condi}) are imposed at the origin of the conformal time. 
We shall see now that they are equivalent to  the Bunch-Davies vacuum conditions.
Since the initial time corresponds
to a large value of redshift, it corresponds to asymptotic times in the past
in a very good approximation. More precisely, the conformal time is related with the redshift $ z $ by
\bea
&& \eta \;= \;\frac{3 \, t_0}{\sqrt{\;1\;+\;z}} \quad : \quad {\rm matter ~ dominated ~ era} \; , \cr \cr
&& \eta \;= \;\frac{2 \, t_0 \; \sqrt{\;1\;+\;z_{\rm eq}}}{1\;+\;z}\;+ \;\frac{t_0}{\sqrt{\;1\; + \;z_{\rm eq}}}
 \quad : \quad {\rm radiation ~ dominated ~ era} \;
\eea
where $ t_0 = 13.7 $ Gyr is the age of the universe, $ 1 + z_{\rm eq} = 3048 $ is the transition
from the radiation dominated to the matter dominated era. $ \eta_0 = 3 \, t_0 $ corresponds 
to the present time. For $ z \gg z_{\rm eq} $ we see that,
$$
\eta \;\simeq \;\frac{t_0}{\sqrt{\,1 + z_{\rm eq}}} \;= \; 0.018 \; t_0 \; .
$$ 
Hence, the conformal time at decoupling differs from the conformal time today $ \eta_0 = 3 \, t_0 $ 
by an amount $ \sim 3 \, t_0 $.  Hence, the initial time can be considered as an asymptotic time
deep in the past. More precisely, the change in the phases of the mode functions
is characterized by $ (M \, t_0)  \sim 3 \times 10^{30} $ for a typical mass $ M \sim 4 $ meV. 
Hence, the initial conditions for the mode functions Eqs.(\ref{ciB}) and (\ref{condi}) are virtually
identical to Bunch-Davies initial conditions.

\bigskip

The vacuum density and energy density Eqs.(\ref{condasi}) and (\ref{hp}) are determined by the 
short distance behaviour of the two point function in coordinate space. 
In momentum space, it is the high energy
behaviour that dominates the vacuum density and energy density for late times.
The physical quantities can be written as integrals of mode functions as in Eqs.(\ref{dce}) 
and (\ref{psiex}).
One can see that the relevant comoving momenta $k$'s contributing at a physical energy scale $ q $
take the value $ k =  q \; a(\eta) $. At late times, (e. g. today) $ a(\eta) \sim z_{\;\rm decoupling} $ and 
therefore only large $ k \sim z_{\rm decoupling} \; M $ are relevant. 
This fact decreases the effect of the initial conditions. 
Analogous effects take place for the initial conditions of inflationary fluctuations with the exception of the low multipoles, particularly the quadrupole \cite{condi}.

\bigskip

\begin{figure}[htbp]
\centering
\includegraphics[width=10cm,height=12.72cm]{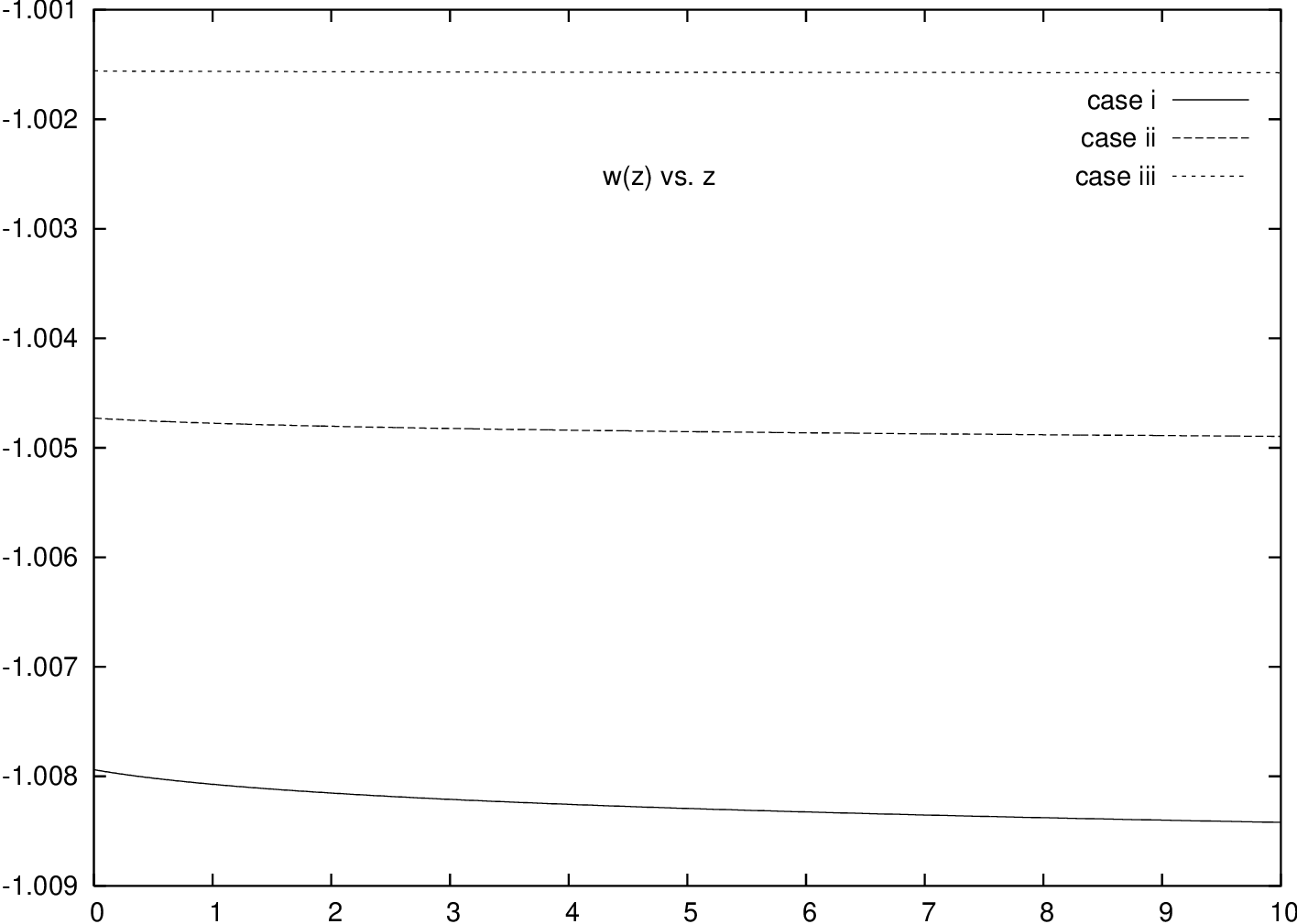}
\caption{ {\it The Equation of State  w(z)\; vs. the redshift z} for the three cases explicitly calculated in this paper : 
{\it(i)} [full line] No neutrino contribution to the dark energy and the scalar mass
$ M = 3.96 $ meV. \;
{\it(ii)} [broken line] A Majorana neutrino with mass $ m = 3.2 $ meV and the scalar mass
$ M = 4.35 $ meV. \;
{\it (iii)} [dotted line] A Dirac neutrino with mass $ m = 3.2 $ meV and the scalar mass
$ M = 4.66 $ meV. [See the discussion in Section  \ref{mM}]. \; In all three cases $ w < - 1 $ by less than
$ 1 \%$. }
\label{wz}
\end{figure}
\begin{itemize}
\item{ In {\it Fig. \ref{wz}}  we plot {\it the equation of state  w(z)  as a function of 
 z } for the three cases explicitly calculated in this paper: } 

\item{(i) No neutrino contribution to the dark energy and the scalar mass
$ M = 3.96 $ meV.}
\item{(ii) A Majorana neutrino with mass $ m = 3.2 $ meV and the scalar mass
$ M = 4.35 $ meV.}
\item{(iii) A Dirac neutrino with mass $ m = 3.2 $ meV and the scalar mass
$ M = 4.66 $ meV. [See the discussion in Section  \ref{mM}].}

\item{We see that the equation of state in all the three cases (i)-(iii) differs from the cosmological
constant case $ w = - 1 $ by less than $ 1 \% $.} 

\item{The value of the lightest neutrino mass Eq.(\ref{mneu}) 
is well below the neutrino mass splittings $ \sqrt{\Delta m^2_{sun}} $ and $ \sqrt{\Delta m^2_{atm}} $
and  consistent with both direct and inverse mass hierarchies. 
A quasi-degenerate mass spectrum will give a large negative contribution to the dark energy
and will require a scalar particle with a mass $ M \gtrsim 100$ meV to reproduce the observed
dark energy data Eq.(\ref{dato}). Such a particle can very well exist 
but it cannot be the axion [see Eq.(\ref{cotma})]. Indeed, the scalar particle can have the mass 
value given by Eq.(\ref{sinu})
in case all three neutrinos decay in a time scale of the age of the universe in order to dissipate
their cosmological vacuum energy as discussed in Section \ref{estneu}.}

\item{On the other hand, a range of neutrino masses
from $10^{-3}$ eV to $0.1$ eV in agreement with neutrino mass differences from oscillations and the value Eq.(\ref{mneu}) for the mass of the lightest neutrinos is compatible with a consistent baryogenesis.}
\end{itemize}

\section{Conclusions}

\begin{itemize}
\item{We find that the presence of a cosmological {\bf quantum vacuum} energy with an equation of state
just below $ - 1 $ is the  {\it unavoidable} consequence of the existence
of light particles with very weak couplings.
Bosons yield positive contributions and fermions yield negative contributions
to the vacuum energy.}

\item{It must be noticed the present lack of knowledge about the {\it low energy} (energy $ \sim 1 $ meV)
particle physics region. Actually, most of the constraints on this sector follow from
astrophysics and cosmology  including the {\it new constraints} that 
we obtain here on the axion mass.}

\item{No exotic physics needs to be invoked to explain the dark energy.
Since the observed energy scale of the dark energy is very low, we find natural to explain it only
through low energy physics. The effects from energy scales higher than
1 eV or even 1 MeV arrive strongly suppressed to the dark energy scale of 1 meV.}
\item {In summary, dark energy can be explained by a very light and very weakly coupled scalar particle
which decouples by redshift $ z_S \gg 1 $. If the scalar particle is the axion, 
then $ z_S \sim 2.2 \times 10^{18} $.

We have four main cases:}

\item{{\bf(i)} No neutrino contribution. This happens when the lightest neutrino has a mass  
 $ m \ll 1 $ meV and when the vacuum neutrino contribution dissipates 
in the time scale of the age of the universe [see Eq.(\ref{autoa})].
The scalar mass must be 
\be\label{sinnuC}
M \;= \;\frac{10.1 \; {\rm meV}}{\left(\; \log z_S\;-\;\frac14\; \right)^{ \! \frac14}}
\quad : \quad {\rm no ~ vacuum ~ neutrino ~ energy } \; .
\ee
If the scalar is the axion, then $ M = 3.96 $ meV in this case.}

\item{{\bf(ii)} The lightest neutrino is Majorana and has a mass $ m \simeq 3.2 $ meV.
Then, the scalar mass must be 
$$
M \;= \;\frac{11.1 \; {\rm meV}}{\left(\; \log z_S \;-\; \frac14 \; \right)^{ \! \frac14}}
\quad : \quad {\rm the ~  Majorana ~ neutrino ~ contributes}
$$
If the scalar is the axion, then $ M = 4.35 $ meV in this case.}

\item{{\bf(iii)} The lightest neutrino is Dirac and has a mass $ m \simeq 3.2 $ meV.
Then, the scalar mass must be 
$$
M \;=\; \frac{11.9 \; {\rm meV}}{\left(\; \log z_S \;- \;\frac14 \;\right)^{ \! \frac14}}
\quad : \quad {\rm the ~ Dirac ~ neutrino ~ contributes}
$$
If the scalar is the axion, then $ M = 4.66 $ meV in this case.}

\item{Therefore, {\it in all the three cases (i)-(iii)} above where the axion explains the dark energy {\bf we predict} 
its mass in the range:
\be\label{cotaxi}
3.96 \; {\rm meV} \;< M \;< \; 4.66 \; {\rm meV} \; .
\ee
The left and right ends of the interval in Eq.(\ref{cotaxi}) correspond to no neutrino contribution and to the lightest neutrino as
a Dirac fermion with mass $ m = 3.2 $ meV, respectively.}

\item{In short, we uncovered here the general mechanism producing the dark energy today. This mechanism
has it grounds on well known quantum physics and cosmology. 
The dark energy appears as a {\bf quantum vacuum effect} arising when stable and weakly coupled
quantum fields live in expanding cosmological space-times. That is to say, dark energy in the universe today is a QFT effect in (classical) curved space-times. That is to say, this is a {\it semiclassical gravity effect}.}

{\item In addition, we have found here that the {\it axion} with mass in the {\it meV range} is a very serious candidate for dark energy, while we have shown already \cite{HdVNSaxion}, \cite{UniversekeV} that it is robustely excluded as a dark matter candidate. The cosmic dark energy  today  is in the  {\it meV scale} while the dark matter (cosmic and galactic) particle is in the keV scale \cite{HdVNSkeV}, \cite{UniversekeV}.}

\item{Many research avenues open now connecting dark energy and light particles physics.
The more immediate being:}

\item{{\bf(1)} The study of the radiative corrections to the axion and 
neutrino cosmological vacuum evolution from their interactions.}
\item{{\bf(2)} The study of the early neutrino and axion dynamics at temperatures $ \gtrsim 1 $ MeV and $ \gtrsim  10^6 $ GeV, respectively.}
\item{{\bf(3)} The study of particle propagation in the media formed by the axion and the neutrino vacuum.}
\item{{\bf(4)} Last but not least:  The probable deep connection between dark energy and dark matter
through low energy particle states beyond the Standard Model of particle physics.}
\end{itemize}

\bigskip

\appendix

\section{Dimensional and Cutoff  Regularization of the Vacuum Energy }\label{regu}

Physical vacuum quantities are computed in Section \ref{coqv} as the equal point limit of
two point functions. The distance $ z $ between the points Eq.(\ref{defz}) naturally plays the role 
of the regularization parameter. Alternatively, one can regularize the two point function with 
dimensional regularization or cutoff regularization and set $ z = 0 $ in the regularized expressions.

In dimensional regularization, we have
\be\label{2ptd}
 G_{\epsilon}(M \, a,a) \;\equiv\; 
<T \; \varphi({\vec x},\eta) \; \varphi({\vec x},\eta)> \;=\; \frac1{a^2} \; 
\int \frac{d^{4- 2 \, \epsilon} k }{(2 \, \pi)^{4- 2 \, \epsilon}} \; \; \frac{i}{k^2 -a^2 \; M^2 +i \, 0}
\ee 
\be\; = \;
\frac{M^{2- 2 \, \epsilon}}{(4 \, \pi)^{2 - \epsilon}} \; \;\frac1{a^{2 \, \epsilon}} \;\; \Gamma(\epsilon - 1)
\; .
\ee
Subtracting the value in Minkowski space-time ($ a=1 $) yields,
\be\label{renG}
G_{\epsilon}(M \, a,a) - G_{\epsilon}(M,1) \;= \; M^{(2- 2 \; \epsilon)} \; \;
\frac{\Gamma(\epsilon - 1)}{(4 \, \pi)^{(2 - \epsilon)}} \;\left[\, a^{-2 \, \epsilon}
- 1 \,\right]\; \buildrel{\epsilon \to 0}\over =  \;\frac{M^2}{2 \, (2 \, \pi)^2} \; \log a
\ee
in agreement with Eq.(\ref{denss}).

\medskip

Alternatively, by regularizing with an ultraviolet cutoff $ \Lambda $ in four space-time dimensions,
we have
\bea\label{2ptc}
 &&G_{\Lambda}(M \, a)\; \equiv \; \frac1{a^2} \; 
\int \frac{d^4 k }{(2 \, \pi)^4} \;  \frac{i}{k^2 -a^2 \; M^2 +i \, 0}\; = \;  
\left( \frac{\Lambda}{4 \, \pi \; a}\right)^2 - \left( \frac{M}{4 \, \pi}\right)^2 \;
\log\left[1+ \left( \frac{\Lambda}{M \; a}\right)^2\right] =\cr \cr
&&\buildrel{\Lambda \to \infty}\over= \left( \frac{\Lambda}{4 \, \pi \; a}\right)^2 -
\frac{M^2}{2 \, (2 \, \pi)^2} \;\log\left[\,\frac{\Lambda}{M \; a}\,\right] \; .
\eea
\;\;

Subtracting the divergence in $ \Lambda = \infty $ again leads to the result Eqs. (\ref{renG}) 
and (\ref{denss}):
\;\;

$$
G_{\Lambda}(M \, a) -  \left[ \left( \frac{\Lambda}{4 \, \pi \; a}\right)^2- 
\frac{M^2}{2 \, (2 \, \pi)^2} \;\log\frac{\Lambda}{M}\right]\; \buildrel{\Lambda \to \infty}\over=\;
\frac{M^2}{2 \, (2 \, \pi)^2} \; \log a \; .$$
\;\;

We have therefore verified that the point splitting regularization used in Section \ref{coqv} as well as
dimensional and cutoff regularization methods yield identical results.
(It is known since longtime that dimensional regularization gives
the same physical results than other regularization methods \cite{lapluta}).
Analogous results are valid for the two point fermion function.

\bigskip
\bigskip

{\bf Acknowledgment:} We thank Daniel Boyanovsky, P. Astier, G. Smoot, S. Perlmutter, B. Schmidt and A. Riess for useful discussions,
and Carlos Frenk for interesting correspondence.

\end{document}